\documentclass[conference]{IEEEtran}
\IEEEoverridecommandlockouts
\usepackage{cite}
\usepackage{amsmath,amssymb,amsfonts}
\usepackage{algorithmic}
\usepackage{graphicx}
\usepackage{multicol}
\usepackage{textcomp}
\usepackage{xcolor}
\usepackage{dsfont}
\usepackage{subcaption}
\usepackage{arydshln}
\def\BibTeX{{\rm B\kern-.05em{\sc i\kern-.025em b}\kern-.08em
    T\kern-.1667em\lower.7ex\hbox{E}\kern-.125emX}}
    
\begin{document}

\title{Clinically Calibrated Machine Learning Benchmarks for Large-Scale Multi-Disorder EEG Classification
}

\author{\IEEEauthorblockN{Argha Kamal Samanta}
\IEEEauthorblockA{
\textit{Department of Electronics and}\\
\textit{Electrical Communication Engineering}\\
\textit{Indian Institute of Technology, Kharagpur}\\
West Bengal, India\\
arghakamal25@gmail.com
}
\and
\IEEEauthorblockN{Deepak Mewada}
\IEEEauthorblockA{
\textit{Department of Computer}\\
\textit{Science and Engineering}\\
\textit{Indian Institute of Technology, Kharagpur}\\
West Bengal, India\\
deepakmewada96@kgpian.iitkgp.ac.in
}
\and
\IEEEauthorblockN{Monalisa Sarma}
\IEEEauthorblockA{
\textit{Subir Chowdhury School of }\\
\textit{Quality and Reliability}\\
\textit{Indian Institute of Technology, Kharagpur}\\
West Bengal, India\\
monalisa@iitkgp.ac.in
}

\and
\IEEEauthorblockN{Debasis Samanta}
\IEEEauthorblockA{
\textit{Department of Computer}\\
\textit{Science and Engineering}\\
\textit{Indian Institute of Technology, Kharagpur}\\
West Bengal, India\\
dsamanta@iitkgp.ac.in
}
}

\maketitle

\begin{abstract}
Clinical electroencephalography is routinely used to evaluate patients with diverse and often overlapping neurological conditions, yet interpretation remains manual, time-intensive, and variable across experts. While automated EEG analysis has been widely studied, most existing methods target isolated diagnostic problems, particularly seizure detection, and provide limited support for multi-disorder clinical screening.

This study examines automated EEG-based classification across eleven clinically relevant neurological disorder categories, encompassing acute time-critical conditions, chronic neurocognitive and developmental disorders, and disorders with indirect or weak electrophysiological signatures. EEG recordings are processed using a standard longitudinal bipolar montage and represented through a multi-domain feature set capturing temporal statistics, spectral structure, signal complexity, and inter-channel relationships. Disorder-aware machine learning models are trained under severe class imbalance, with decision thresholds explicitly calibrated to prioritize diagnostic sensitivity.

Evaluation on a large, heterogeneous clinical EEG dataset demonstrates that sensitivity-oriented modeling achieves recall exceeding 80\% for the majority of disorder categories, with several low-prevalence conditions showing absolute recall gains of 15–30\% after threshold calibration compared to default operating points. Feature importance analysis reveals physiologically plausible patterns consistent with established clinical EEG markers.

These results establish realistic performance baselines for multi-disorder EEG classification and provide quantitative evidence that sensitivity-prioritized automated analysis can support scalable EEG screening and triage in real-world clinical settings.
\end{abstract}


\section{Introduction}

Electroencephalography (EEG) is one of the most frequently ordered neurodiagnostic investigations worldwide, with an estimated 20--30 million clinical EEG studies performed annually across outpatient, inpatient, and intensive care settings \cite{Tatum2014, Westover2015}. Its use extends far beyond seizure evaluation. EEG is routinely requested to assess altered mental status, encephalopathy, developmental delay, cognitive decline, suspected cerebrovascular events, movement disorders, sleep disturbances, and peripheral nervous system dysfunction, often in patients with overlapping or uncertain diagnoses \cite{Schomer2018,Benbadis2008}. In these contexts, EEG serves primarily as a \emph{screening and triage tool}, helping clinicians narrow differential diagnoses and prioritize further investigation or intervention.

However, the scale and complexity of modern EEG practice have outpaced the capacity of manual interpretation. Continuous EEG monitoring in intensive care units commonly generates 24--72 hours of multichannel data per patient, while long-term ambulatory recordings frequently exceed 24 hours \cite{Hirsch2013}. At the same time, the number of trained clinical neurophysiologists has grown far more slowly than EEG utilization, leading to substantial reporting delays and increased diagnostic variability \cite{Halford2017}. Inter-rater agreement for clinically important EEG findings has been reported to range from moderate to poor for several pattern classes, particularly outside of overt seizures \cite{Benbadis2008, Westover2015}. These constraints create a practical risk: clinically relevant abnormalities may be missed, detected late, or interpreted inconsistently.

Automated EEG analysis has therefore become an active area of research. Most prior work has focused on seizure detection or seizure type classification, motivated by the high clinical urgency of epilepsy and the availability of labeled datasets \cite{Shoeb2010,  Roy2019}. While these efforts have demonstrated that machine learning models can detect stereotyped EEG events under controlled conditions, their scope remains narrow. Large portions of routine clinical EEG practice involve disorders with subtler, distributed, or indirect electrophysiological signatures, for which automated support remains limited \cite{Jeong2004, Babiloni2016}. Moreover, the prevailing single-disorder paradigm does not reflect clinical reality, where EEG is rarely ordered to answer a single binary question.

The neurological disorders considered in this study were selected to reflect three major clinical EEG use-cases that together account for a substantial fraction of EEG utilization and diagnostic burden. First, acute and time-sensitive conditions such as seizure disorders and cerebrovascular diseases are associated with high morbidity and mortality if detection is delayed. Nonconvulsive seizures occur in up to 20--30\% of critically ill patients with altered consciousness, yet are frequently underdiagnosed without continuous EEG monitoring \cite{Hirsch2013, Westover2015}. Cerebrovascular diseases remain a leading cause of death and long-term disability worldwide, with EEG often used to evaluate unexplained encephalopathy, post-stroke seizures, and secondary brain dysfunction \cite{Feigin2019}.

Second, chronic neurodevelopmental and neurodegenerative conditions, including developmental delay, cerebral degeneration, and behavioral or cognitive syndromes, represent a growing public health challenge. Dementia affects more than 55 million individuals globally, with prevalence expected to double by 2050 \cite{WHO2021Dementia}. EEG abnormalities in these disorders are typically diffuse and evolve slowly, making visual interpretation subjective and prone to inter-observer disagreement \cite{Jeong2004, Babiloni2016}. Similarly, EEG evaluation in developmental delay and cognitive syndromes often relies on qualitative pattern recognition across long recordings, despite evidence that subtle spectral and complexity changes carry diagnostic information \cite{Dauwels2010}.

Third, several clinically relevant disorders present indirect or weak EEG correlates. Peripheral nervous system disorders, movement and cerebellar disorders, and headache disorders are common reasons for EEG referral, yet their electrophysiological signatures are often non-specific and easily overlooked \cite{Nuwer2018, Cassim2001}. Migraine alone affects more than one billion individuals worldwide and is a leading cause of disability, yet EEG findings remain inconsistently interpreted and rarely supported by automated tools \cite{GBD2019Migraine}. In these cases, the value of EEG lies not in definitive diagnosis, but in supporting exclusion, stratification, and clinical decision-making.

Taken together, these disorder groups span distinct electrophysiological mechanisms, levels of diagnostic urgency, and prevalence profiles. Addressing them simultaneously reflects actual EEG referral patterns more faithfully than disorder-specific models. Automated classification across multiple neurological conditions is therefore not a cosmetic extension of existing methods, but a structurally different and more demanding problem. It requires models that can operate under severe class imbalance, heterogeneous signal quality, and overlapping clinical phenotypes, while prioritizing sensitivity for conditions where missed detection carries significant risk \cite{Fawcett2006, Saito2015}.

Recent availability of large-scale clinical EEG repositories has made systematic investigation of this problem feasible. The Harvard Electroencephalography Database (HEEDB) aggregates hundreds of thousands of EEG recordings acquired in routine clinical environments and linked with diagnostic metadata \cite{Sun2025, Zafar2025}. Unlike curated research datasets, HEEDB captures the full variability of clinical EEG, including long recording durations, artifacts, comorbid conditions, and highly imbalanced diagnostic labels. Despite its scale, comprehensive machine learning baselines for multi-disorder EEG classification on this dataset remain largely absent.

This work addresses this gap by establishing clinically grounded machine learning baselines for the simultaneous classification of eleven neurological disorder categories from EEG. The proposed framework emphasizes diagnostic sensitivity and calibrated decision thresholds rather than accuracy alone, aligning model behavior with clinical priorities. By integrating standard bipolar montaging, multi-domain electrophysiological feature extraction, and disorder-aware modeling strategies, this study aims to quantify what is realistically achievable for automated EEG-based screening and triage under real-world clinical conditions.

The contributions of this work are fourfold: (i) systematic evaluation of automated EEG classification across eleven clinically meaningful neurological disorders within a unified framework; (ii) establishment of realistic performance baselines on a large, heterogeneous clinical EEG dataset; (iii) demonstration of sensitivity-prioritized decision modeling for low-prevalence and high-risk conditions; and (iv) disorder-specific insights into the relationship between electrophysiological signatures and model behavior. These results provide a foundation for future development of scalable, clinically integrated EEG decision support systems.

The remainder of this paper is organized as follows. Section~II reviews related work on automated EEG analysis and large-scale clinical EEG datasets. Section~III describes the dataset and background considerations. Section~IV details the proposed methodology. Section~V presents experimental results. Section~VI discusses clinical implications and limitations, followed by conclusions in Section~VII.

\section{Literature Survey}

\subsection{Automated EEG-Based Neurological Disorder Classification}

The application of machine learning and deep learning to EEG-based neurological disorder classification has advanced significantly in recent years. A systematic review of studies from 2013 to August 2024 identified epilepsy, depression, and Alzheimer's disease as the most studied conditions, with 32, 12, and 10 studies respectively \cite{article}. Electroencephalogram serves as an essential tool for exploring brain activity and holds particular importance in mental health research, enabling detection of various conditions including epilepsy, schizophrenia, bipolar disorder, major depressive disorder, ADHD, anxiety disorders, sleep disorders, and neurodevelopmental disorders \cite{Rahul2024EEGSchizophreniaReview}.

Recent studies have demonstrated remarkable classification accuracy, with some approaches achieving near-perfect performance: wearable depression detection devices using CNN-LSTM models reached 99.9\% accuracy, quantum-based machine learning with QSVM achieved 100\% success in schizophrenia detection, and the Adazd-Net model attained 99.85\% accuracy in Alzheimer's diagnosis \cite{article}. EEG data classification plays a pivotal role in diagnosing epilepsy through recognition of abnormal patterns such as sharp wave discharges, assessing sleep disorders by analyzing distinct patterns across different sleep stages, evaluating brain injuries including traumatic brain injury and stroke, and diagnosing neurological disorders such as Parkinson's disease and multiple sclerosis \cite{10.1145/3742795}.

Despite promising results, automated feature extraction from EEG signals remains a challenging task, particularly for neuropsychiatric disorders which demand extraction of neuro-markers for use in automated classification \cite{PARSA2023107683}. Traditional machine learning methods often rely on manual feature engineering, which is demanding and time-consuming, motivating the development of deep learning approaches that can automatically learn and extract meaningful representations from raw data \cite{Shams2020NeurologicalDisorderDL}.

\subsection{Deep Learning Architectures for EEG Analysis}

Deep learning-based cross-dimensional spatiotemporal joint representation learning has become a key enabler for EEG-based neurological disorder classification, though most existing approaches either process spatial and temporal information separately or adopt simplistic concatenation-based fusion strategies that fail to capture intrinsic cross-dimensional interactions \cite{DONG2025108982}. Recent work on epileptic seizure classification introduced attention-based deep convolutional neural networks that combine multi-head self-attention mechanisms with CNNs to classify seven subtypes of generalized and focal epileptic seizures \cite{GILL2024109732}.

Time-frequency transformations have been employed to enhance visualization and interpretation of EEG data, with studies converting EEG signals into time-frequency images using novel transforms to improve accuracy of visual inspection and deep learning-based classification \cite{Amer2024VisualSignalNeuroscience}. Motor imagery EEG signal classification faces challenges including signal noise, inter-subject variability, and real-time processing demands, with recent approaches combining empirical mode decomposition for extracting intrinsic signal modes and continuous wavelet transform for multi-resolution analysis, achieving 95.7\% accuracy on benchmark datasets \cite{Mathiyazhagan2025MotorImageryEEG}.

Transformer-based models have emerged for EEG classification, with architectures incorporating convolutional stems for input embedding and time-frequency multi-head cross-attention for integrating time-domain patterns into frequency points, demonstrating superiority over traditional methods \cite{ZEYNALI2023105130}. However, RNN-based models suffer from limitations when dealing with high-dimensional, multi-channel EEG signals, with their channel-independent design ignoring neurophysiological coupling among leads, while Transformer-based models treat spatial EEG data as flattened sequences, disregarding neuroanatomical constraints and losing spatial adjacency information \cite{DONG2025108982}.

\subsection{Gradient Boosting and XGBoost for EEG Classification}

Extreme Gradient Boosting (XGBoost) has demonstrated efficacy in EEG-based classification tasks, with studies achieving 99.1\% accuracy in emotion classification using statistical features extracted from sliding time windows \cite{Khamthung2024EmotionEEGXGBoost}, and showing significantly improved modeling efficiency compared to general gradient boosting decision trees. XGBoost has been successfully applied to motor imagery classification using Common Spatial Patterns features, eliminating the need for frequency band selection and demonstrating robustness to random noise \cite{9342132}.

Hybrid approaches combining CNN feature extraction with XGBoost classification have achieved exceptional results, with CNN-XGBoost fusion methods obtaining 99.712\% accuracy for arousal, 99.770\% for valence, and 99.770\% for dominance in emotion recognition from EEG spectrogram images \cite{Khan2022CNNXGBoostEEG}. For epileptic seizure detection, XGBoost combined with complementary ensemble empirical mode decomposition (CEEMD) has demonstrated superior performance by decomposing raw EEG signals into components and extracting multi-domain features, with the fast implementation of XGBoost allowing rapid multiple iterations and providing interpretability through feature importance measures \cite{e22020140, Vanabelle2019SeizureXGBoost}.

To address severe class imbalance in medical datasets, ensemble approaches combining multiple XGBoost models through bagging have been developed, linking individual classifiers in parallel and splitting training data into balanced subsets through random under-sampling or over-sampling, demonstrating effectiveness for disorders of consciousness detection despite heavy dataset skew \cite{WANG2022116778}.

\subsection{Multi-Layer Perceptron and Neural Networks for EEG}

Multilayer perceptrons provide a straightforward approach for capturing global patterns in EEG data, and when appropriately designed with optimal numbers of neuron layers and neurons per layer, can act as universal approximators, though overtraining problems may arise due to low signal-to-noise ratio of EEG signals \cite{10.3389/fnins.2025.1541062}. Compared to traditional machine learning algorithms which produced 88\% accuracy with 11 seconds of computation time, subject-independent generalized MLP models have successfully classified motor imagery signals with approximately 90\% accuracy and half the classification time, demonstrating greater efficacy with bio-signals \cite{SHARMA2022103101}.

MLPs have been widely used for classifying EEG signals with high efficiency, particularly for pattern classification tasks such as determining EEG signal patterns produced during different motor movements, employing activation functions like tan-sigmoid in hidden and output layers \cite{Hamzah2016EEGMotorMLP}. For intracranial EEG epileptic seizure recognition, optimally configured MLPs with tan-sigmoid transfer functions for input-hidden layers and pure linear functions for hidden-output layers, trained using Levenberg-Marquardt algorithms, have demonstrated effective classification performance \cite{RAGHU2017205}.

MLPs have achieved 69.69\% average classification accuracy across five subjects for four emotional categories using hemispheric asymmetry alpha power indices, significantly exceeding chance probability \cite{4428831}. For brain-computer interface neurofeedback applications, MLP neural networks have demonstrated accurate detection of motor tasks versus rest using movement-related cortical potential features, with performance comparable to prior literature while requiring minimal preprocessing steps \cite{9896584}.

\subsection{Large-Scale Clinical EEG Databases}

The Harvard Electroencephalography Database (HEEDB) represents a comprehensive clinical resource, aggregating more than 280,000 EEG recordings from more than 108,000 patients across four Harvard-affiliated hospitals, harmonized using the Brain Imaging Data Structure and linked with clinical notes, ICD-10 codes, medications, and EEG reports \cite{Sun2025, Zafar2025}. HEEDB includes routine, epilepsy monitoring unit, and intensive care unit EEGs across all age groups, with 73\% linked to deidentified clinical reports and 96\% of those matched to recordings, following HIPAA Safe Harbor deidentification standards \cite{Sun2025, Zafar2025}.

The Temple University Hospital EEG Corpus represents another major publicly available resource, containing clinical EEG data collected over 14 years with curated records paired with textual clinician reports describing patients' clinical conditions \cite{10.3389/fnins.2016.00196}. Historical EEG signal processing tools have been devised using either ad hoc heuristic methods or training pattern recognition engines on small datasets, yielding limited results due to the great variability in brain signals that can only be properly interpreted by building statistical models using massive amounts of data \cite{10.3389/fnins.2016.00196}.

Despite EEG being perhaps the most pervasive modality for acquiring brain signals, there has been a severe lack of publicly available data, with most existing databases containing limited recordings: the EEG Motor Movement/Imagery Dataset contains approximately 1500 recordings from 109 subjects, the CHB-MIT database includes data from 22 subjects mostly pediatric, and the Karunya University database contains 175 16-channel EEGs of 10-second duration \cite{10.3389/fnins.2016.00196}. HEEDB fills a critical gap in EEG data availability for epilepsy research, enabling large-scale, privacy-compliant, and clinically relevant analysis that accelerates development of diagnostic tools and improves training datasets for machine learning \cite{Sun2025, Zafar2025}.

\subsection{Research Gap and Motivation}

Despite significant advances in automated EEG classification, several critical gaps remain. Most existing studies evaluate models on relatively small, homogeneous datasets that fail to capture the complexity and variability of real-world clinical populations. The lack of massive amounts of publicly available EEG data has prevented the application of state-of-the-art machine learning algorithms to discover new diagnostics and validate clinical practice \cite{10.3389/fnins.2016.00196}. Furthermore, while individual disorder classification has been extensively studied, comprehensive multi-disorder benchmarking on large-scale heterogeneous datasets remains limited.

There has been a notable increase in publications applying AI techniques to EEG-based disorder detection over the past two years, with a particularly sharp rise in 2024, attributed to advancements in deep learning technologies, growing interest in non-invasive diagnostic tools, and improved computational resources \cite{article}. However, most research focuses on achieving high accuracy rather than clinical sensitivity, overlooking the reality that false negatives in medical diagnosis carry significantly higher costs than false positives.

The HEEDB dataset, despite its scale and clinical richness, currently lacks comprehensive diagnostic baselines across its diverse neurological pathology annotations. With more than 280,000 EEG recordings accompanied by metadata, clinical notes, and physician interpretations, HEEDB represents an actively maintained and continuously expanding resource \cite{Sun2025, Zafar2025}, yet systematic evaluation of machine learning approaches across its full spectrum of disorders has not been conducted. Additionally, the severe class imbalance inherent in clinical populations, where rare but critical conditions may represent less than 5\% of cases, requires specialized approaches beyond standard accuracy optimization.

This work addresses these gaps by establishing the first comprehensive diagnostic baselines for 11 distinct neurological conditions within HEEDB, employing a biologically-informed approach that matches computational architectures to specific electrophysiological signatures, and prioritizing clinical sensitivity through threshold optimization to ensure reliable detection even for low-prevalence disorders.

\section{Methodology}
\label{sec:methodology}

This section formulates the proposed EEG-based multi-disorder classification framework in mathematical terms. Each processing stage corresponds directly to the blocks shown in Fig.~\ref{fig:system_overview}, ensuring a one-to-one mapping between the conceptual framework and its algorithmic realization.

\subsection{Problem Definition}

A clinical EEG recording is represented as
\[
\mathbf{X} \in \mathbb{R}^{C \times T},
\]
where \(C\) denotes the number of channels and \(T\) the number of temporal samples acquired at sampling frequency \(f_s\). Each recording is associated with a binary label vector
\[
\mathbf{y} = [y_1,\dots,y_{11}]^\top \in \{0,1\}^{11},
\]
where \(y_k=1\) indicates the presence of the \(k\)-th neurological disorder.

The objective is to estimate disorder-specific posterior probabilities
\[
f_k(\mathbf{X}) = P(y_k=1 \mid \mathbf{X}), \quad k=1,\dots,11,
\]
with emphasis on recall due to the asymmetric clinical cost of false negatives. The severe class imbalance motivating this choice is summarized in Table~\ref{tab:class_distribution}.

\subsection{Pipeline Overview}

\begin{figure*}[htbp]
    \centering
    \includegraphics[width=0.99\textwidth]{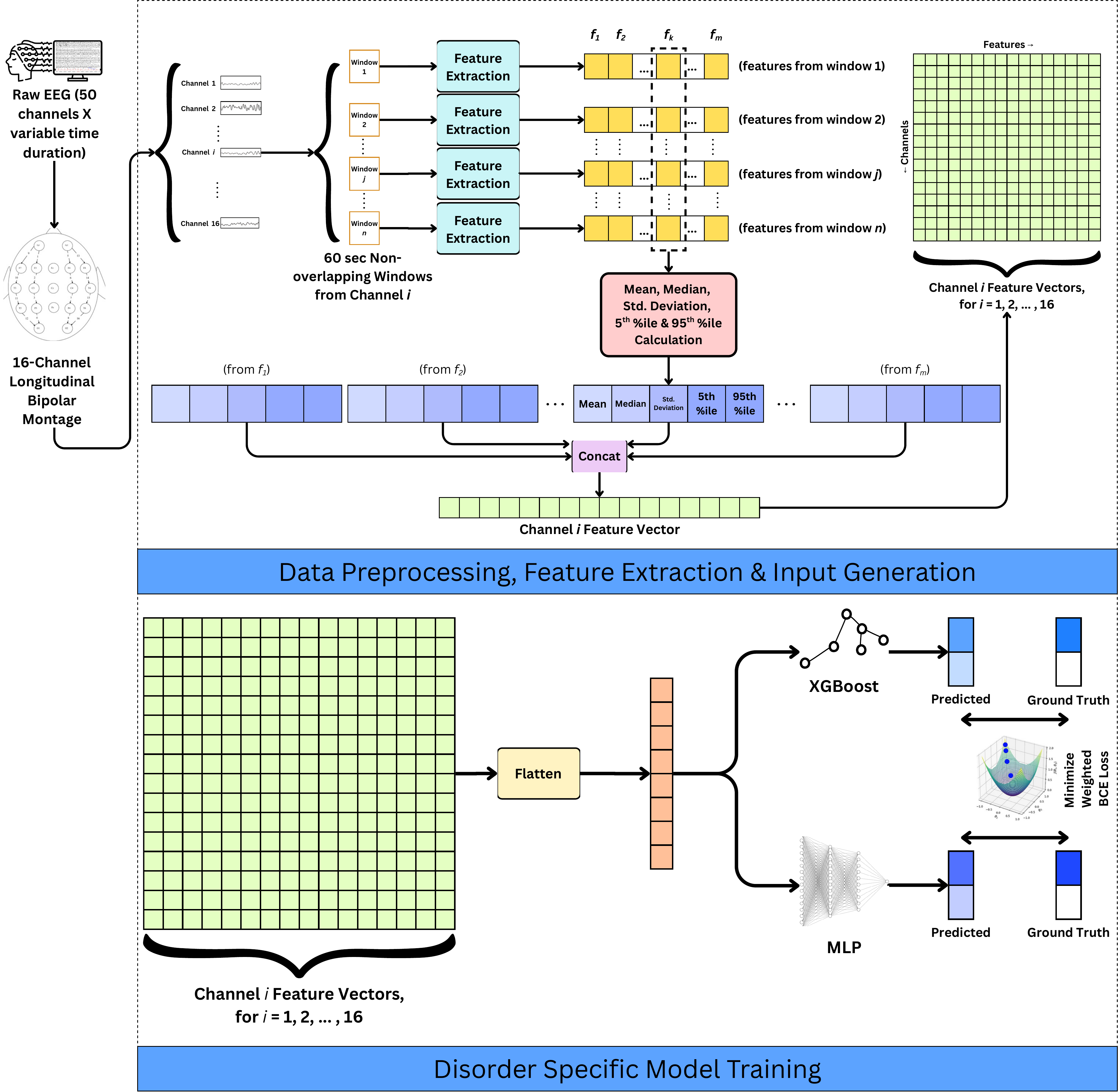}
    \caption{Proposed end-to-end framework for EEG-based multi-disorder classification.}
    \label{fig:system_overview}
\end{figure*}

The complete processing pipeline is illustrated in Fig.~\ref{fig:system_overview}. Raw EEG signals undergo spatial referencing, temporal segmentation, multi-domain feature extraction, statistical aggregation, and disorder-specific classification with sensitivity-oriented threshold calibration.

\subsection{Spatial Referencing}

\begin{figure}[htbp]
    \centering
    \includegraphics[width=0.48\textwidth]{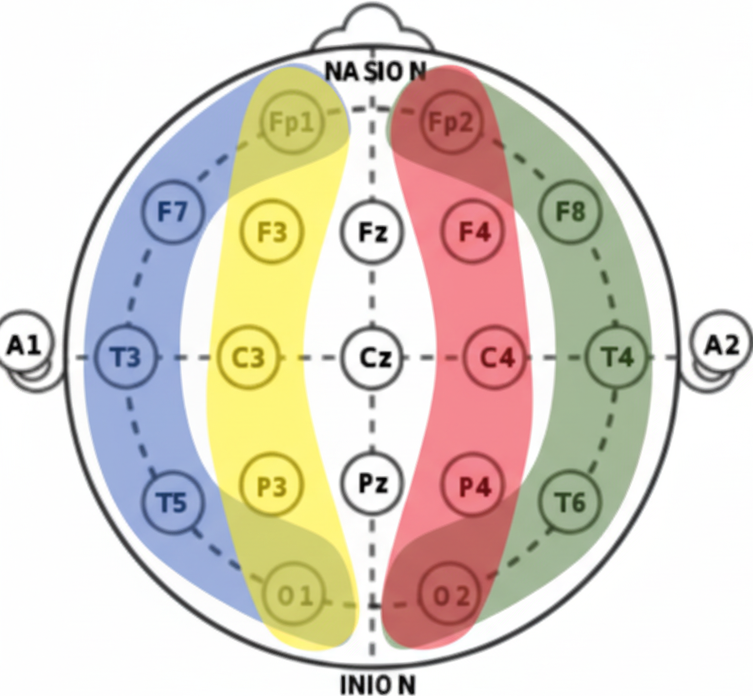}
    \caption{
    \textbf{Illustration of the ACNS longitudinal bipolar montage (“double-banana” configuration).}  
    Electrode pairs are grouped into four principal anatomical chains spanning both hemispheres:  
    {blue line-}{ Left Lateral (LL) Chain} (Fp1–F7, F7–T3, T3–T5, T5–O1),  
    {yellow line-}{ Left Parasagittal (LP) Chain} (Fp1–F3, F3–C3, C3–P3, P3–O1),  
    {red line-}{Right Parasagittal (RP) Chain} (Fp2–F4, F4–C4, C4–P4, P4–O2), and  
    {green line-}{Right Lateral (RL) Chain} (Fp2–F8, F8–T4, T4–T6, T6–O2).  
     } 
    \label{fig:bipolar_montage}
\end{figure}

EEG signals are transformed into a 16-channel longitudinal bipolar montage derived from the international 10--20 system \cite{Klem1999TenTwentySystem,Baker2008EEGMCI}. Let \(\mathbf{X}(t)\in\mathbb{R}^C\) denote the channel vector at time \(t\). Bipolar referencing is expressed as
\[
\mathbf{B}(t) = \mathbf{D}\mathbf{X}(t),
\]
where \(\mathbf{D}\in\mathbb{R}^{C_b\times C}\) is a sparse differencing matrix encoding adjacent electrode pairs (Fig.~\ref{fig:bipolar_montage}). This linear transformation suppresses common-mode noise while emphasizing local spatial gradients.

\subsection{Temporal Segmentation}

Each bipolar channel is segmented into non-overlapping windows of duration \(\Delta t = 60\) s:
\[
\mathbf{B}_i \rightarrow \{\mathbf{b}_i^{(w)}\}_{w=1}^{W}, \quad
\mathbf{b}_i^{(w)}\in\mathbb{R}^{N_w}.
\]
This assumes approximate wide-sense stationarity within windows, enabling reliable estimation of statistical and spectral quantities.

\subsection{Multi-Domain Feature Extraction}

Feature extraction is formulated as a nonlinear mapping
\[
\Phi:\mathbb{R}^{N_w}\rightarrow\mathbb{R}^d,
\]
applied independently to each window and channel. After aggregation, 3,108 features are obtained per recording.

\subsubsection{Time-Domain Statistics}

Statistical moments up to fourth order are computed:
\[
\mu,\ \sigma^2,\ 
\gamma_1=\frac{\mathbb{E}[(b-\mu)^3]}{\sigma^3},\ 
\gamma_2=\frac{\mathbb{E}[(b-\mu)^4]}{\sigma^4}-3.
\]
Line length,
\[
L=\sum_{n=1}^{N_w-1}|b[n+1]-b[n]|,
\]
approximates total variation and is sensitive to sharp transients.

\subsubsection{Hjorth Parameters and Entropy}

Hjorth activity, mobility, and complexity \cite{hjorth1970eeg} summarize variance and frequency dispersion. Signal unpredictability is quantified using Shannon entropy \cite{6773024},
\[
H=-\sum_{k=1}^{K}p_k\log_2 p_k.
\]

\subsubsection{Spectral Features}

Power spectral density \(S(f)\) is estimated using Welch’s method \cite{Welch1967PowerSpectra}. Relative band power is computed as
\[
P_{\beta,\text{rel}}=
\frac{\int_{\beta}S(f)\,df}{\int_{0.5}^{45}S(f)\,df}.
\]

\subsubsection{Functional Connectivity}

Inter-channel synchronization is quantified using Pearson correlation \cite{10.1098/rspl.1895.0041},
\[
\rho_{xy}=\frac{\mathrm{cov}(x,y)}{\sigma_x\sigma_y}.
\]

\subsection{Feature Aggregation}

Window-level features are aggregated using mean, standard deviation, and percentile statistics, yielding a recording-level representation \(\mathbf{z}\). Median imputation is applied for missing values, followed by z-score normalization.

\subsection{Classification Models}

\begin{figure*}[htbp]
    \centering
    \includegraphics[width=0.48\textwidth]{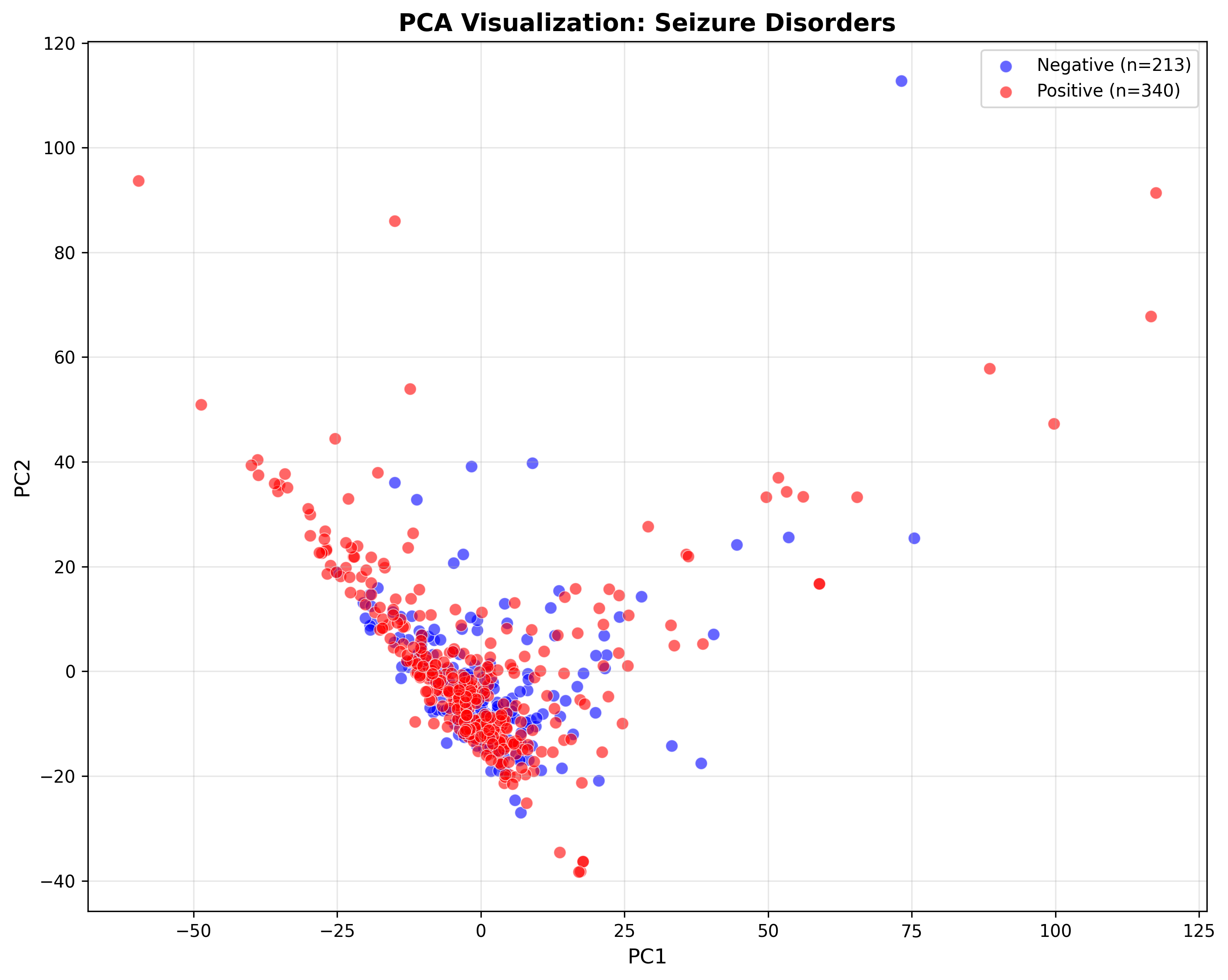}
    \hfill
    \includegraphics[width=0.48\textwidth]{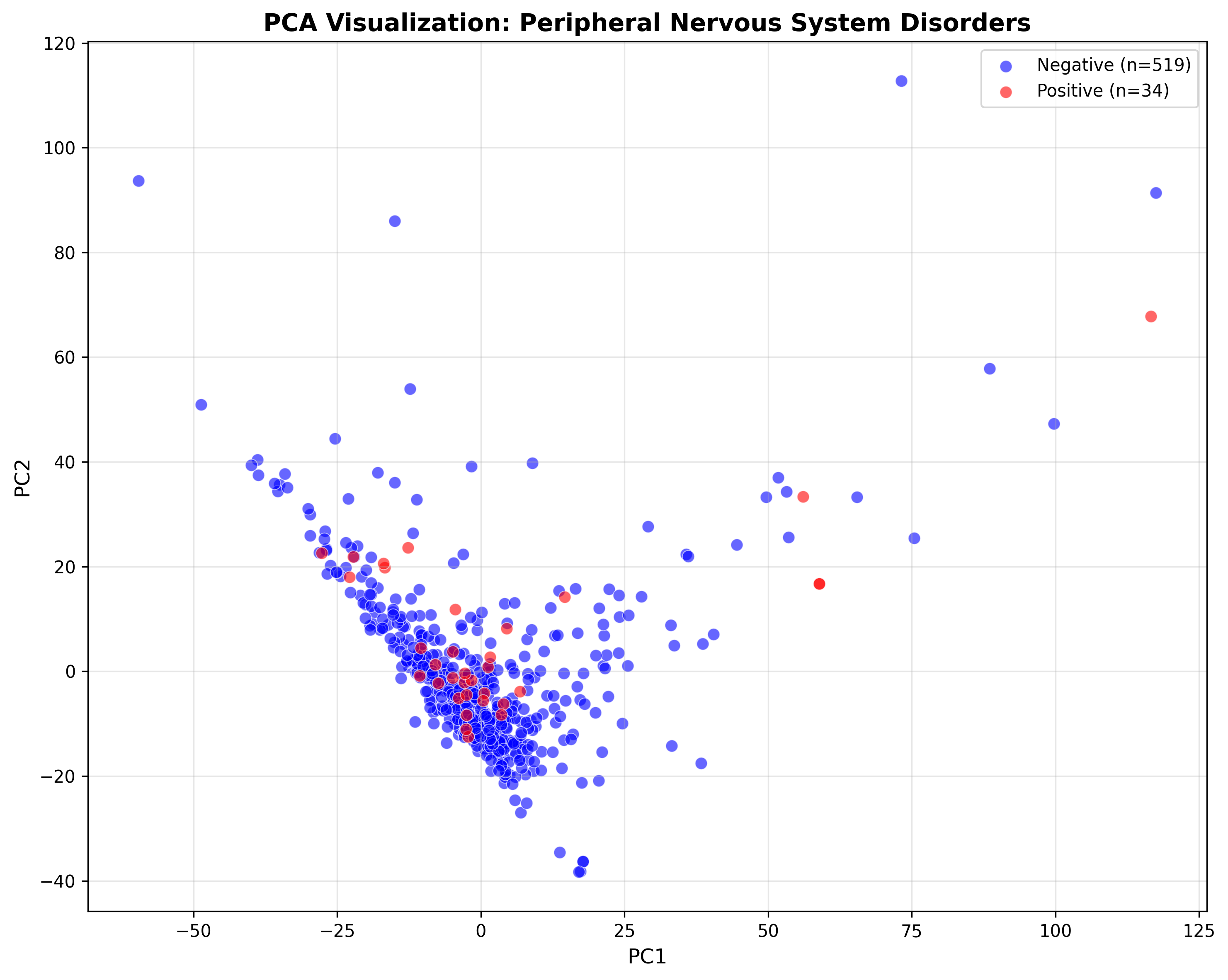}
    \caption{PCA visualizations for seizure-related and peripheral nervous system disorders.}
    \label{fig:pca}
\end{figure*}

Disorder-specific classifiers are selected based on feature geometry, visualized using PCA \cite{Pearson01111901} in Fig.~\ref{fig:pca}.

Gradient-boosted decision trees implemented via XGBoost \cite{Chen_2016} minimize a regularized empirical risk. Hyperparameters are reported in Table~\ref{tab:xgboost_params}.

\begin{table}[htbp]
\centering
\caption{XGBoost hyperparameter configuration.}
\label{tab:xgboost_params}
\begin{tabular}{ll}
\hline
Parameter & Value \\
\hline
Max depth & 6 \\
Learning rate & 0.05 \\
Estimators & 300 \\
Subsample & 0.8 \\
Feature subsample & 0.8 \\
$\gamma$ & 0.1 \\
$\alpha$ & 0.1 \\
$\lambda$ & 1.5 \\
\hline
\end{tabular}
\end{table}

For disorders characterized by subtle distributed patterns, a multi-layer perceptron is used. The network employs ReLU activations \cite{10.5555/3104322.3104425} and is optimized using Adam \cite{kingma2017adammethodstochasticoptimization}. Training parameters are listed in Table~\ref{tab:mlp_params}.

\begin{table}[htbp]
\centering
\caption{MLP training configuration.}
\label{tab:mlp_params}
\begin{tabular}{ll}
\hline
Parameter & Value \\
\hline
Hidden layers & (256, 128, 64) \\
Optimizer & Adam \\
Learning rate & 0.001 \\
Batch size & 32 \\
L2 regularization & 0.0001 \\
Early stopping & 20 epochs \\
\hline
\end{tabular}
\end{table}

\subsection{Decision Threshold Optimization}

Predictions are obtained via
\[
\hat{y}_k=\mathbb{I}[f_k(\mathbf{X})\ge\tau_k],
\]
where thresholds \(\tau_k\) are selected to satisfy recall constraints.

\section{Experimental Setup and Results}
\label{sec:experiments}

This section presents a detailed experimental evaluation of the proposed framework on a large-scale clinical EEG dataset. The analysis is structured to characterize the dataset, define experimental objectives and evaluation protocol, and analyze results across disorder categories, model architectures, and decision strategies. All reported results are interpreted in the context of clinical relevance, class imbalance, and electrophysiological plausibility.

\subsection{Dataset Description}

\subsubsection{Harvard Electroencephalography Database (HEEDB)}

Experiments are conducted using the Harvard Electroencephalography Database (HEEDB), focusing on recordings from the I0002-coded hospital. The dataset consists of routine clinical EEG recordings acquired in real-world hospital environments, capturing substantial variability in recording duration, signal quality, and patient demographics.

Table~\ref{tab:dataset_stats} summarizes the dataset characteristics. Of the 844 scanned EEG files, 553 sessions satisfied channel completeness criteria and were retained for analysis, corresponding to 191 unique patients. Each session was processed using a 16-channel longitudinal bipolar montage, and 3,108 features were extracted per recording.

\begin{table}[htbp]
\centering
\caption{Summary statistics of the HEEDB I0002 dataset used in this study. The dataset reflects real-world clinical EEG recordings with long durations and heterogeneous patient populations.}
\label{tab:dataset_stats}
\begin{tabular}{lr}
\hline
Characteristic & Value \\
\hline
Total EDF files scanned & 844 \\
Valid files (complete channels) & 553 \\
Unique patients & 191 \\
Processed sessions & 553 \\
Bipolar channels per recording & 16 \\
Sampling frequency & 256 Hz \\
Features per sample & 3,108 \\
\hline
\end{tabular}
\end{table}

\subsubsection{Clinical Annotation and Disorder Categories}

Clinical diagnoses are derived from ICD-10 codes and mapped to 18 neurological disorder categories. Each patient-session is associated with binary labels indicating the presence or absence of each disorder, resulting in a multi-label classification setting. Eleven disorders contained sufficient positive samples for reliable evaluation and are included in this study. The class distribution for these disorders is summarized in Table~\ref{tab:class_distribution}, revealing severe imbalance ratios ranging from 0.63:1 to 16.84:1.

\subsubsection{Evaluation and Class Imbalance}

\begin{figure}[htbp]
    \centering
    \includegraphics[width=0.5\textwidth]{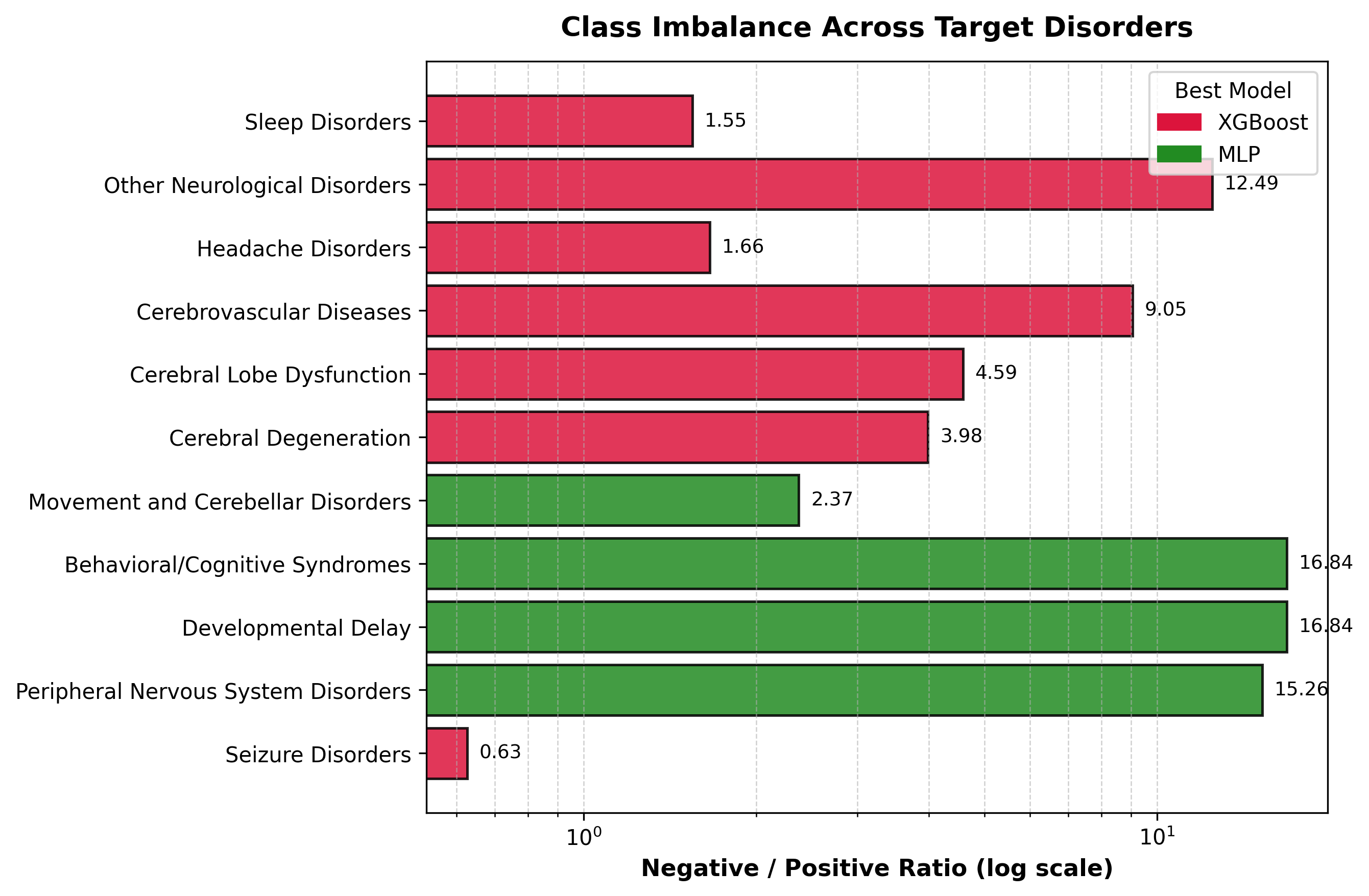}
    \caption{Class imbalance across target disorders.}
    \label{fig:class_imb}
\end{figure}

Evaluation uses patient-level splits with cross-validation. Class imbalance is addressed via weighted losses and threshold calibration. Class distributions are illustrated in Fig.~\ref{fig:class_imb} and summarized in Table~\ref{tab:class_distribution}.

\begin{table}[htbp]
\centering
\caption{Class distribution across target disorders.}
\label{tab:class_distribution}
\begin{tabular}{lrrr}
\hline
Disorder & Positive & Negative & Neg/Pos \\
\hline
Seizure Disorders & 340 & 213 & 0.63 \\
Peripheral NS Disorders & 34 & 519 & 15.26 \\
Developmental Delay & 31 & 522 & 16.84 \\
Behavioral/Cognitive Syndromes & 31 & 522 & 16.84 \\
Movement/Cerebellar Disorders & 164 & 389 & 2.37 \\
Cerebral Degeneration & 111 & 442 & 3.98 \\
Cerebral Lobe Dysfunction & 99 & 454 & 4.59 \\
Cerebrovascular Diseases & 55 & 498 & 9.05 \\
Headache Disorders & 208 & 345 & 1.66 \\
Other Neurological Disorders & 41 & 512 & 12.49 \\
Sleep Disorders & 217 & 336 & 1.55 \\
\hline
\end{tabular}
\end{table}

\subsection{Experimental Objectives}

The experimental evaluation is designed to address the following objectives:

\begin{enumerate}
    \item Establish baseline classification performance across multiple neurological disorders in HEEDB.
    \item Validate whether recall-optimized models can achieve clinically relevant sensitivity thresholds ($\geq$80\%).
    \item Assess the discriminative capacity of multi-domain EEG features across disorder categories.
    \item Compare gradient boosting (XGBoost) and neural network (MLP) architectures under varying class imbalance conditions.
    \item Evaluate robustness to severe class imbalance through weighted losses and threshold optimization.
    \item Verify performance stability under cross-validation.
\end{enumerate}

\subsection{Evaluation Protocol}

\subsubsection{Data Partitioning}

Patient-level train–test splitting is employed to prevent subject-level information leakage, with 80\% of patients used for training and 20\% held out for testing. The training set is further divided into training and validation subsets (80/20 split) for early stopping and hyperparameter tuning.

\subsubsection{Cross-Validation}

Five-fold stratified cross-validation is performed on the training set. Mean and standard deviation of ROC-AUC across folds are monitored to assess performance stability and generalization.

\subsubsection{Performance Metrics}

Models are evaluated using accuracy, precision, recall (sensitivity), F1-score, ROC-AUC, and average precision (PR-AUC). Given the clinical cost of missed diagnoses, recall is prioritized, with decision thresholds optimized accordingly.

\subsection{Results: XGBoost Classification}

\subsubsection{Seizure Disorders}

Seizure disorders represent the most clinically urgent category and exhibit a relatively balanced class distribution compared to other disorders. Table~\ref{tab:seizure_results} reports detailed performance metrics obtained using XGBoost.

\begin{table}[htbp]
\centering
\caption{Classification performance for Seizure Disorders using XGBoost. The optimized threshold prioritizes sensitivity while maintaining acceptable overall accuracy.}
\label{tab:seizure_results}
\begin{tabular}{lc}
\hline
Metric & Value \\
\hline
Optimal threshold & 0.43 \\
Accuracy & 80.2\% \\
Precision (Negative) & 81.8\% \\
Recall (Negative) & 62.8\% \\
Precision (Positive) & 79.5\% \\
Recall (Positive) & \textbf{91.2\%} \\
F1-Score (Positive) & 84.9\% \\
ROC-AUC & 85.9\% \\
Average Precision & 91.6\% \\
\hline
\end{tabular}
\end{table}

\begin{figure}[htbp]
\centering
\includegraphics[width=0.48\textwidth]{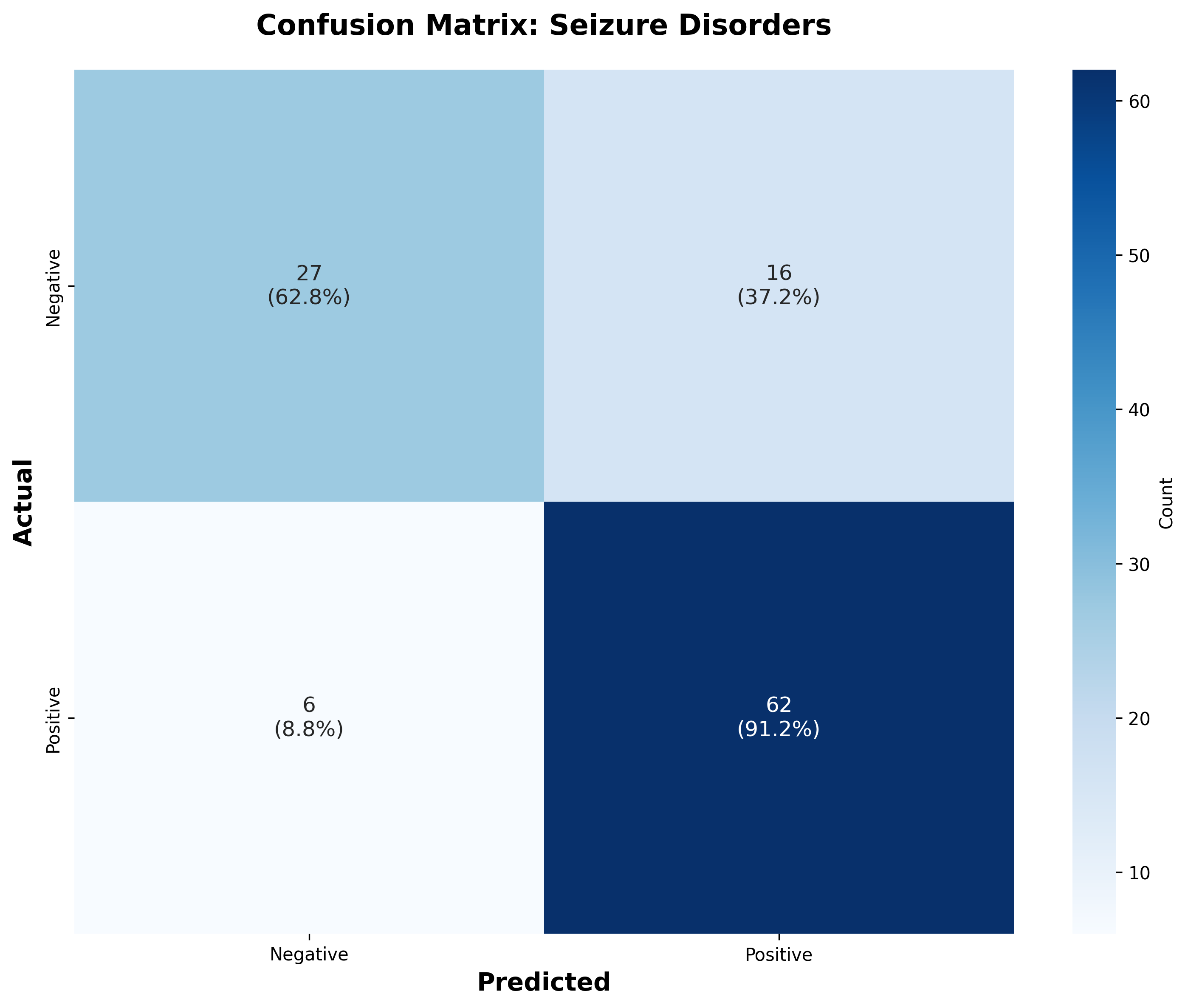}
\caption{Confusion matrix for Seizure Disorders using XGBoost. High sensitivity (91.2\%) is achieved while preserving balanced overall accuracy.}
\label{fig:seizure_cm}
\end{figure}

The confusion matrix in Fig.~\ref{fig:seizure_cm} shows that 91.2\% of seizure-positive cases are correctly identified. This exceeds the predefined clinical sensitivity target of 80\% while maintaining 80.2\% overall accuracy, indicating that threshold optimization effectively shifts the decision boundary toward minimizing false negatives.

\subsubsection{Comprehensive XGBoost Results Across Disorders}

Table~\ref{tab:xgboost_comprehensive} summarizes XGBoost performance across seven neurological disorders.

\begin{table*}[htbp]
\centering
\caption{XGBoost classification performance across seven neurological disorders. Disorder-specific thresholds are optimized to prioritize recall.}
\label{tab:xgboost_comprehensive}
\begin{tabular}{lcccccc}
\hline
Disorder & Threshold & Accuracy & Recall & Avg. Precision & F1 & ROC-AUC \\
\hline
Seizure Disorders & 0.43 & 80.2\% & 91.2\% & 91.6\% & 84.9\% & 85.9\% \\
Cerebral Degeneration & 0.13 & 68.5\% & 81.8\% & 67.8\% & 50.7\% & 82.2\% \\
Cerebral Lobe Dysfunction & 0.10 & 78.4\% & 75.0\% & 72.9\% & 55.6\% & 84.0\% \\
Cerebrovascular Diseases & 0.18 & 69.4\% & 81.8\% & 35.8\% & 34.6\% & 81.5\% \\
Headache Disorders & 0.28 & 76.6\% & 81.0\% & 82.3\% & 72.3\% & 83.7\% \\
Other Neurological Disorders & 0.40 & 75.7\% & 87.5\% & 62.1\% & 34.1\% & 88.6\% \\
Sleep Disorders & 0.25 & 70.3\% & 84.1\% & 76.6\% & 69.2\% & 78.4\% \\
\hline
\end{tabular}
\end{table*}

\begin{figure}[htbp]
\centering
\includegraphics[width=0.48\textwidth]{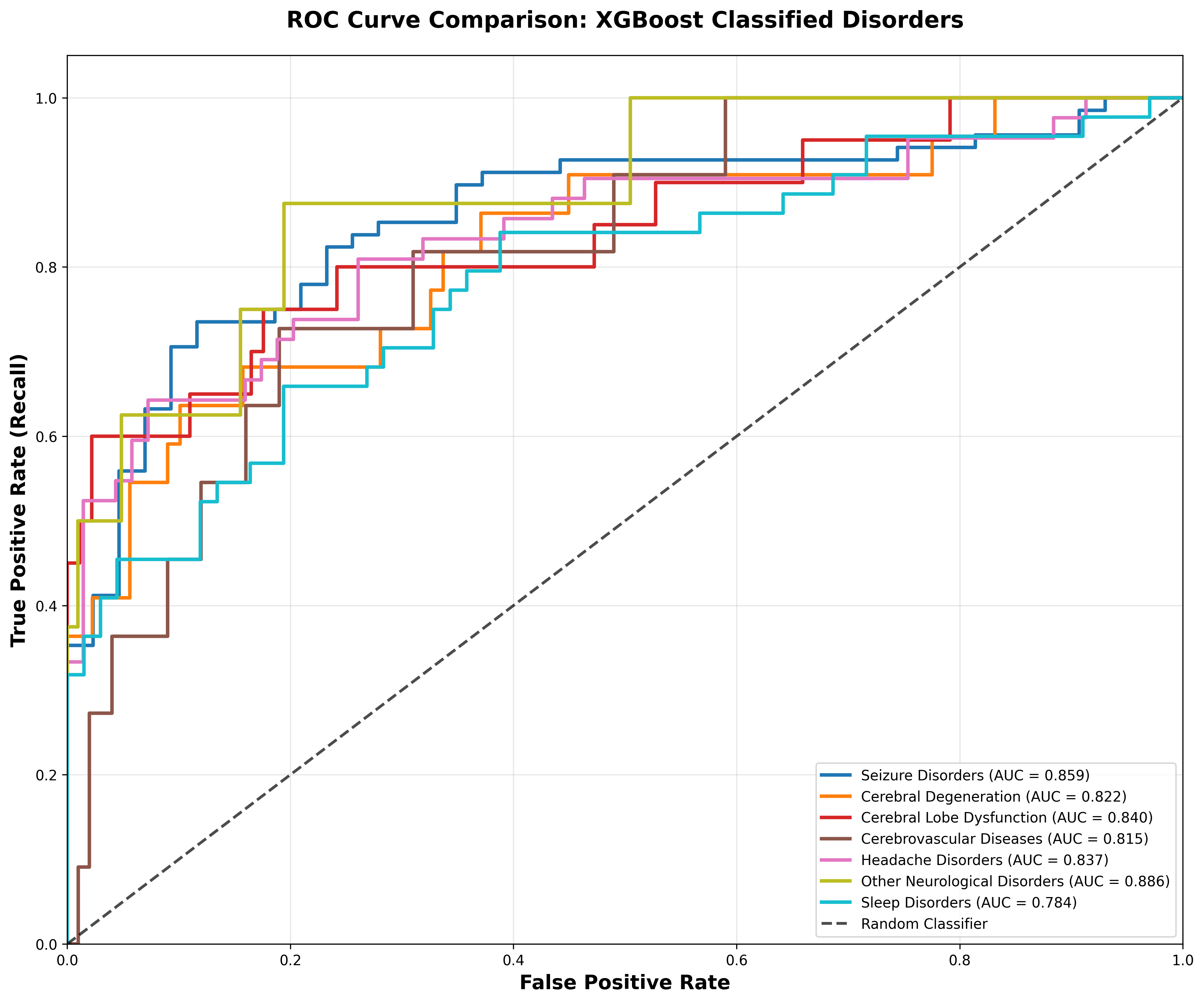}
\caption{ROC curves for XGBoost across seven neurological disorders, indicating consistent discriminative performance.}
\label{fig:xgboost_roc}
\end{figure}

\begin{figure}[htbp]
\centering
\includegraphics[width=0.48\textwidth]{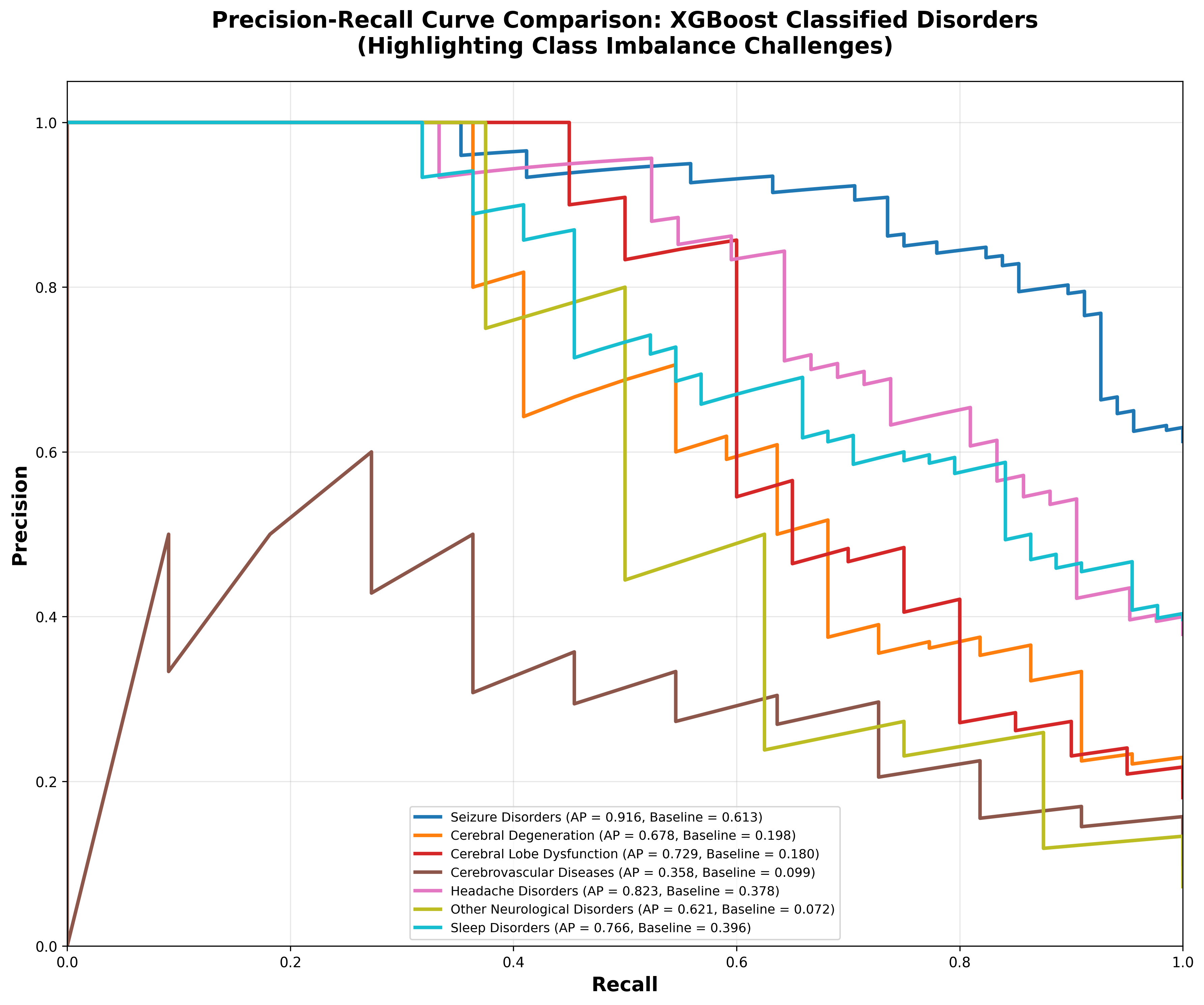}
\caption{Precision–Recall curves for XGBoost, highlighting the impact of class imbalance on precision at high recall levels.}
\label{fig:xgboost_pr}
\end{figure}

\begin{figure}[htbp]
\centering
\includegraphics[width=0.48\textwidth]{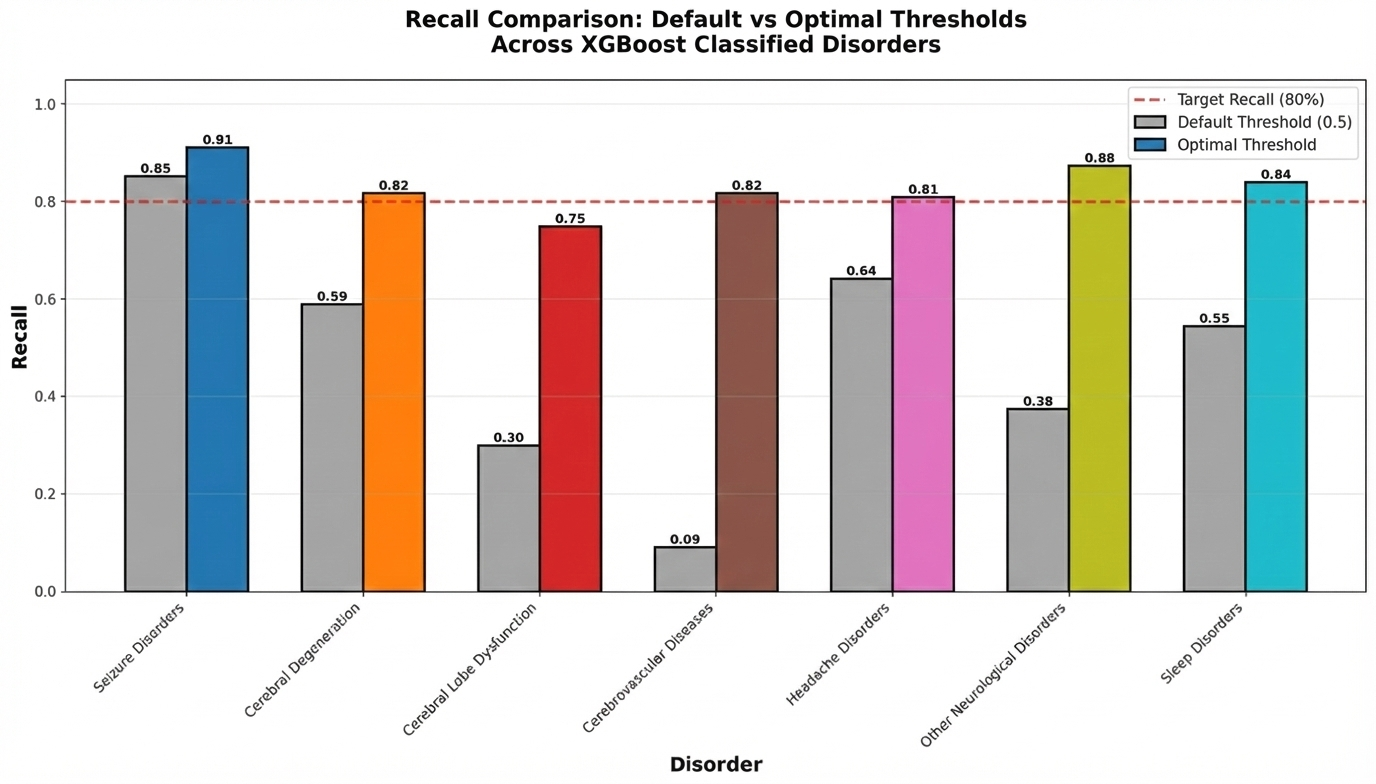}
\caption{Recall achieved by XGBoost across seven neurological disorders. Most disorders meet or exceed the 80\% clinical sensitivity target.}
\label{fig:xgboost_recall}
\end{figure}

ROC curves (Fig.~\ref{fig:xgboost_roc}) and precision–recall curves (Fig.~\ref{fig:xgboost_pr}) show consistent discrimination across disorders, with ROC-AUC values between 78.4\% and 88.6\%. Fig.~\ref{fig:xgboost_recall} further demonstrates that recall exceeds 75\% for all disorders, with six achieving $\geq$80\%.

\subsection{Results: MLP Classification}

\subsubsection{Peripheral Nervous System Disorders}

Table~\ref{tab:peripheral_results} reports MLP performance for Peripheral Nervous System Disorders.

\begin{table}[htbp]
\centering
\caption{MLP performance for Peripheral Nervous System Disorders. High sensitivity is achieved despite severe class imbalance.}
\label{tab:peripheral_results}
\begin{tabular}{lc}
\hline
Metric & Value \\
\hline
Optimal threshold & 0.13 \\
Accuracy & 99.1\% \\
Recall (Positive) & 85.7\% \\
Precision (Positive) & 100\% \\
ROC-AUC & 98.5\% \\
Average Precision & 99\% \\
\hline
\end{tabular}
\end{table}

\begin{figure}[htbp]
\centering
\includegraphics[width=0.48\textwidth]{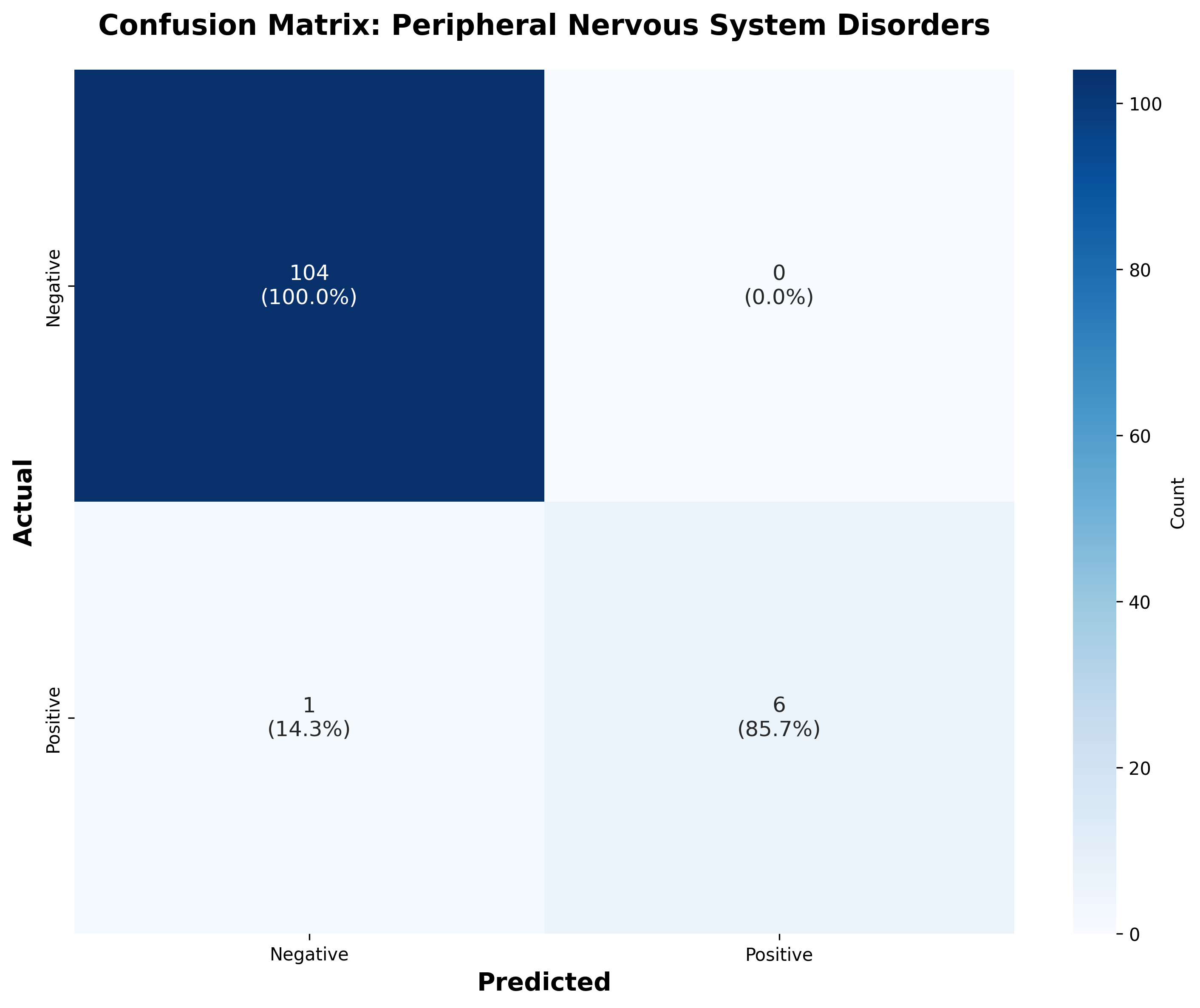}
\caption{Confusion matrix for Peripheral Nervous System Disorders using the MLP model.}
\label{fig:pns_cm}
\end{figure}

\subsubsection{Comprehensive MLP Results}

MLP performance across four neurological disorders is summarized in Table~\ref{tab:mlp_comprehensive}.

\begin{table*}[htbp]
\centering
\caption{MLP classification performance across neurological disorders with diffuse EEG signatures.}
\label{tab:mlp_comprehensive}
\begin{tabular}{lcccccc}
\hline
Disorder & Threshold & Accuracy & Recall & Avg. Precision & F1 & ROC-AUC \\
\hline
Peripheral NS Disorders & 0.13 & 99.1\% & 85.7\% & 91.3\% & 99.0\% & 98.5\% \\
Developmental Delay & 0.03 & 94.6\% & 83.3\% & 50.9\% & 95.0\% & 93.9\% \\
Behavioral/Cognitive Syndromes & 0.06 & 92.8\% & 83.3\% & 71.7\% & 94.0\% & 88.6\% \\
Movement and Cerebellar Disorders & 0.24 & 87.4\% & 81.2\% & 81.3\% & 89.0\% & 90.1\% \\
\hline
\end{tabular}
\end{table*}

\begin{figure}[htbp]
\centering
\includegraphics[width=0.48\textwidth]{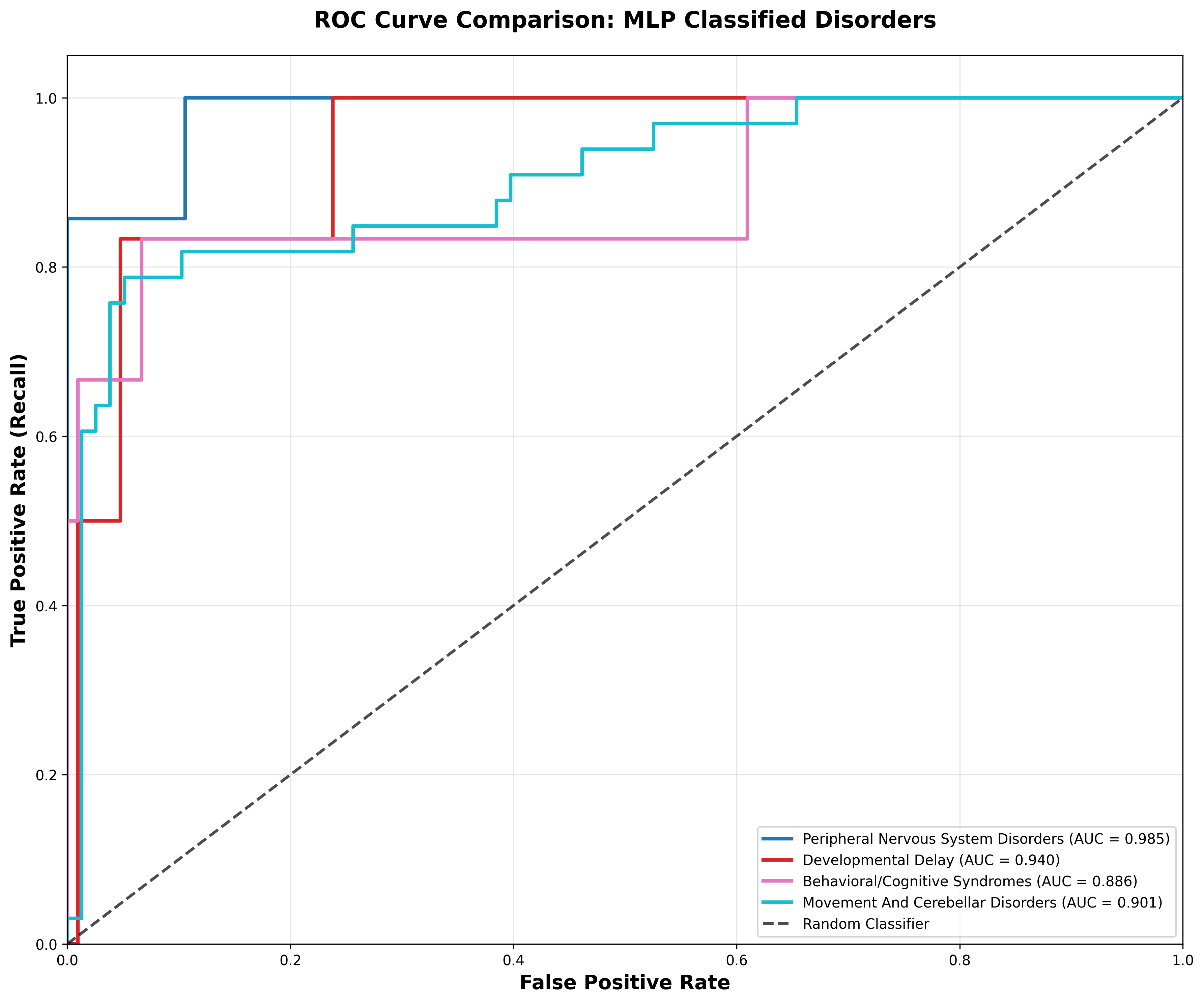}
\caption{ROC curves for the MLP model across four neurological disorders.}
\label{fig:mlp_roc}
\end{figure}

\begin{figure}[htbp]
\centering
\includegraphics[width=0.48\textwidth]{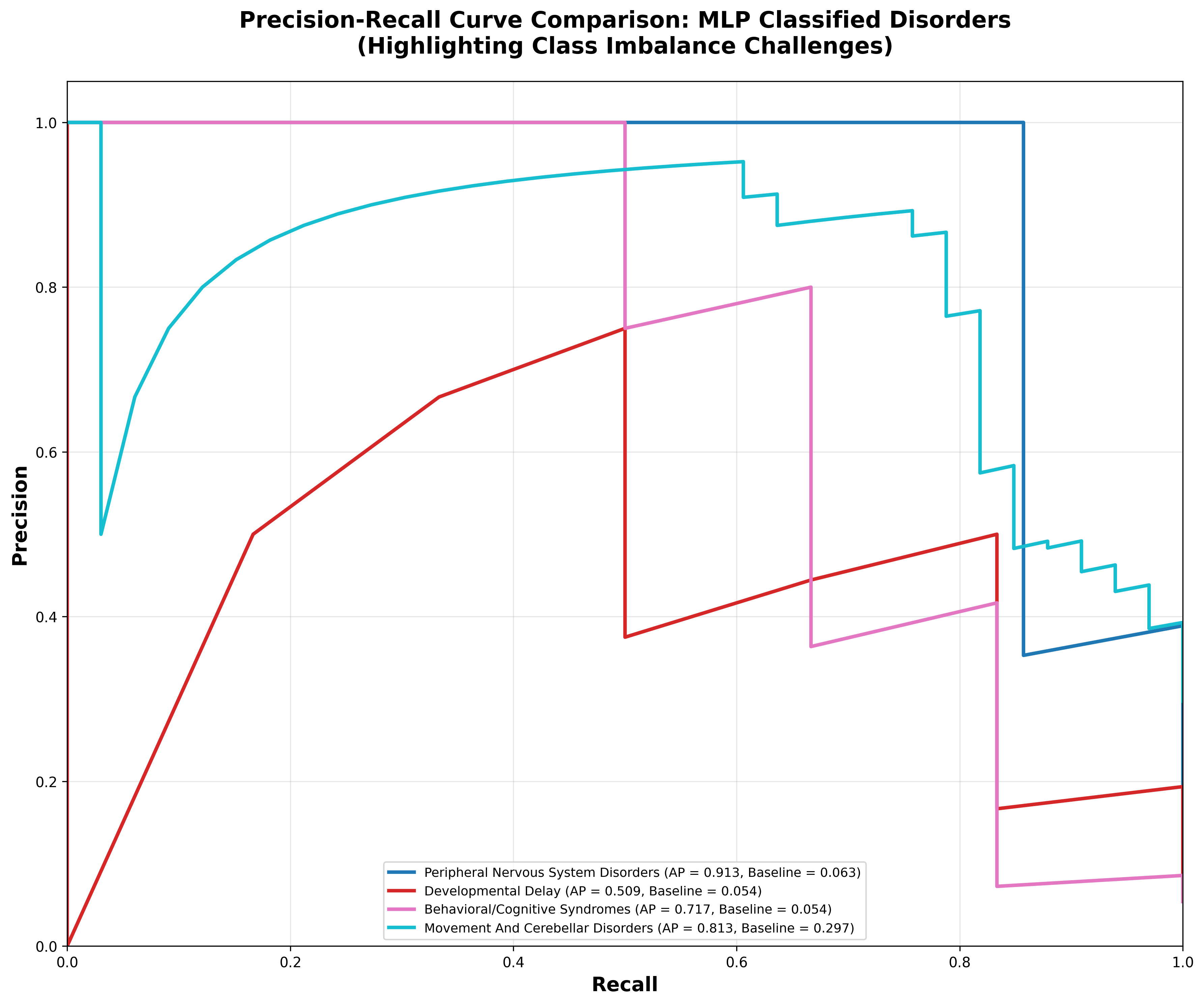}
\caption{Precision–Recall curves for the MLP model under varying class imbalance conditions.}
\label{fig:mlp_pr}
\end{figure}

\begin{figure}[htbp]
\centering
\includegraphics[width=0.48\textwidth]{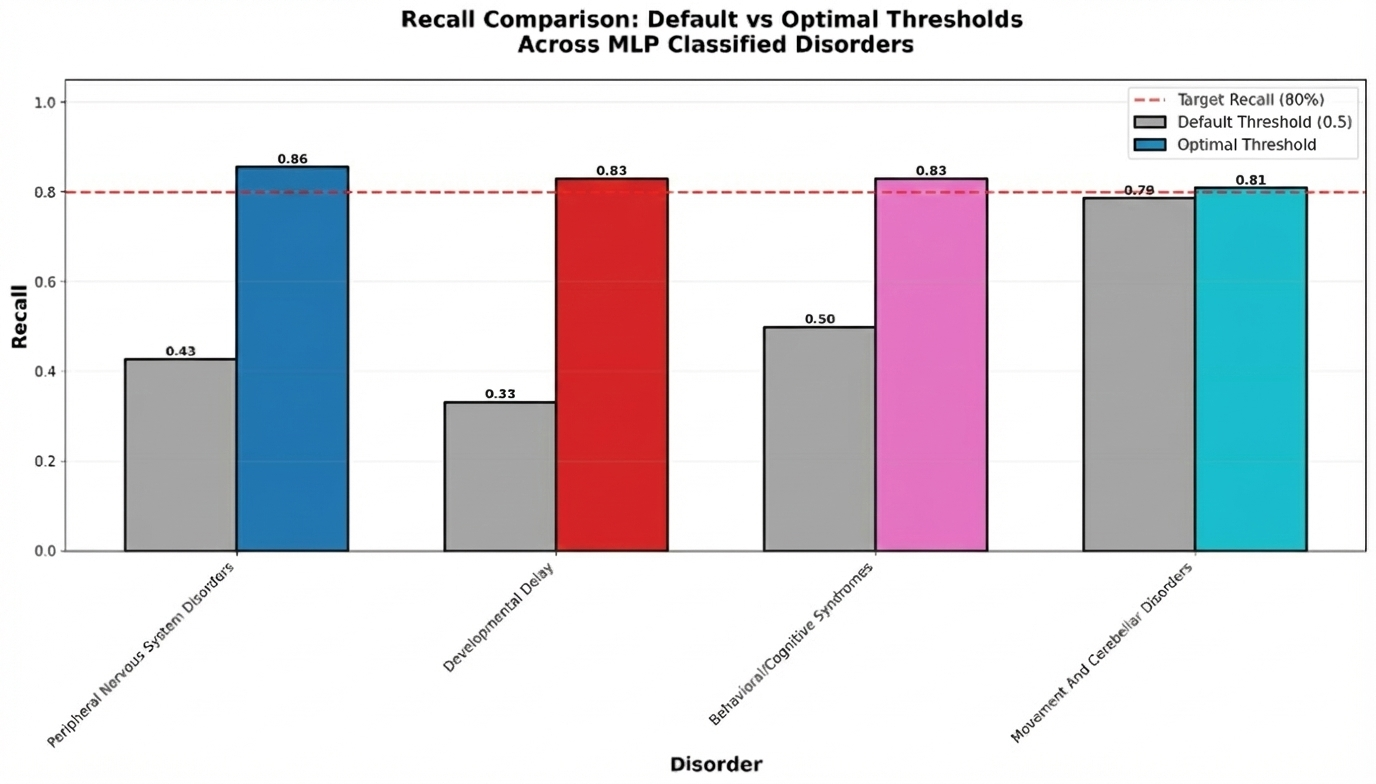}
\caption{Recall achieved by the MLP model across four neurological disorders.}
\label{fig:mlp_recall}
\end{figure}

\subsection{Model Architecture Comparison}

Table~\ref{tab:model_comparison} compares XGBoost and MLP performance across comparable imbalance ratios.

\begin{table*}[htbp]
\centering
\caption{Comparison of XGBoost and MLP performance across disorders with similar imbalance ratios.}
\label{tab:model_comparison}
\begin{tabular}{llccc}
\hline
Neg/Pos Ratio & Disorder & Model & Accuracy & ROC-AUC \\
\hline
1.66:1 & Headache Disorders & XGBoost & 76.6\% & 83.7\% \\
2.37:1 & Movement/Cerebellar & MLP & 87.4\% & 90.1\% \\
3.98:1 & Cerebral Degeneration & XGBoost & 68.5\% & 82.2\% \\
\hdashline
12.49:1 & Other Neurological & XGBoost & 75.7\% & 88.6\% \\
15.26:1 & Peripheral NS & MLP & 99.1\% & 98.5\% \\
16.84:1 & Developmental Delay & MLP & 94.6\% & 93.9\% \\
\hline
\end{tabular}
\end{table*}

\subsection{Feature Importance Analysis}

\subsubsection{Seizure Disorders}

\begin{figure}[htbp]
\centering
\includegraphics[width=0.48\textwidth]{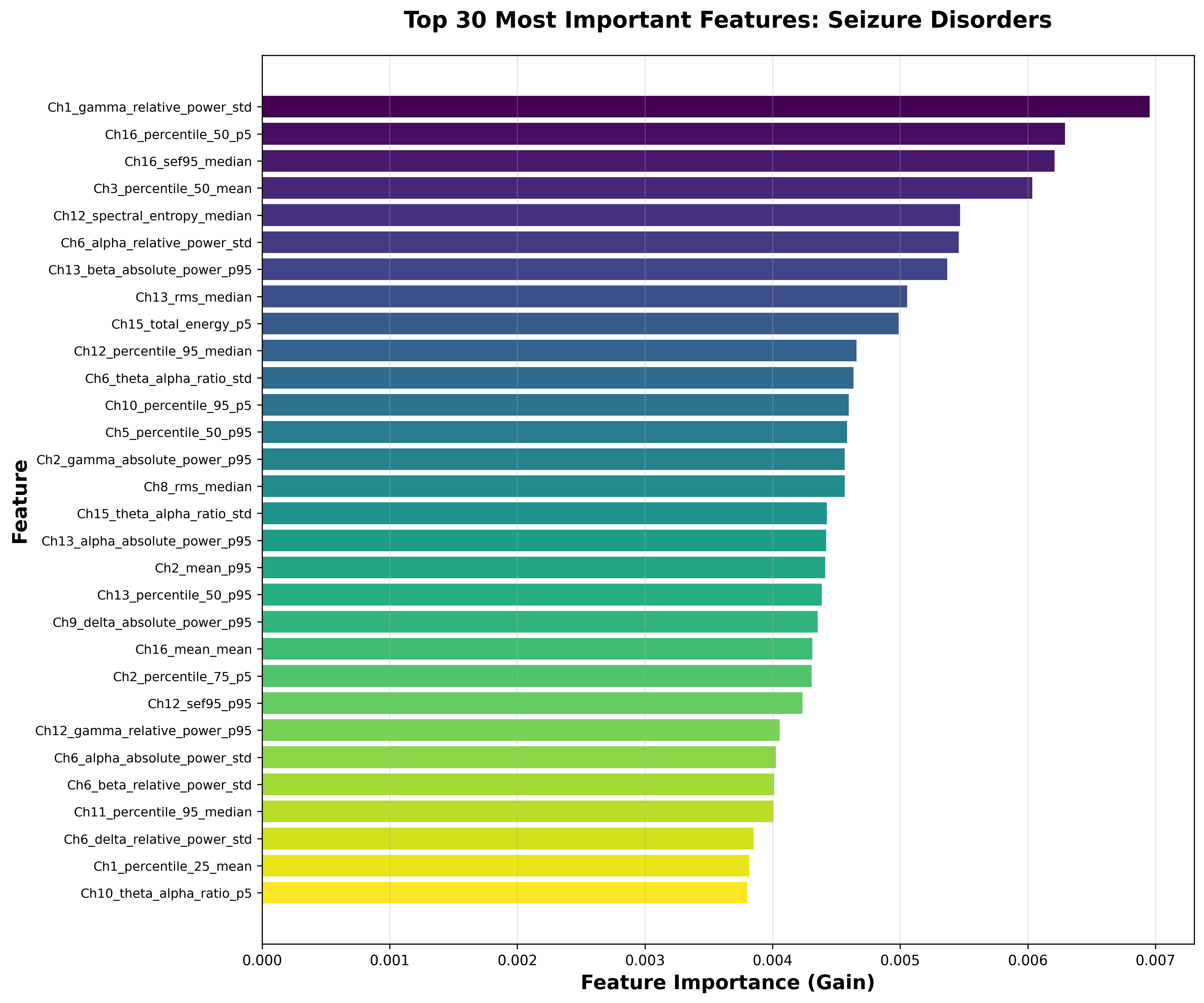}
\caption{Top 30 most informative features for Seizure Disorders. Spectral and energy-based features dominate, consistent with epileptiform EEG activity.}
\label{fig:feature_importance_seizure}
\end{figure}

Table~\ref{tab:feature_importance_seizure} and Fig.~\ref{fig:feature_importance_seizure} summarize the most informative features for seizure classification.

\begin{table}[htbp]
\centering
\caption{Top ten most important features for Seizure Disorders based on XGBoost gain.}
\label{tab:feature_importance_seizure}
\begin{tabular}{llc}
\hline
Rank & Feature & Importance \\
\hline
1 & Ch1\_gamma\_relative\_power\_std & 0.0070 \\
2 & Ch16\_percentile\_50\_p5 & 0.0063 \\
3 & Ch16\_sef95\_median & 0.0062 \\
4 & Ch3\_percentile\_50\_mean & 0.0060 \\
5 & Ch12\_spectral\_entropy\_median & 0.0055 \\
6 & Ch6\_alpha\_relative\_power\_std & 0.0055 \\
7 & Ch13\_beta\_absolute\_power\_p95 & 0.0054 \\
8 & Ch13\_rms\_median & 0.0051 \\
9 & Ch15\_total\_energy\_p5 & 0.0050 \\
10 & Ch12\_percentile\_95\_median & 0.0047 \\
\hline
\end{tabular}
\end{table}

\subsubsection{Peripheral Nervous System Disorders}

\begin{figure}[htbp]
\centering
\includegraphics[width=0.48\textwidth]{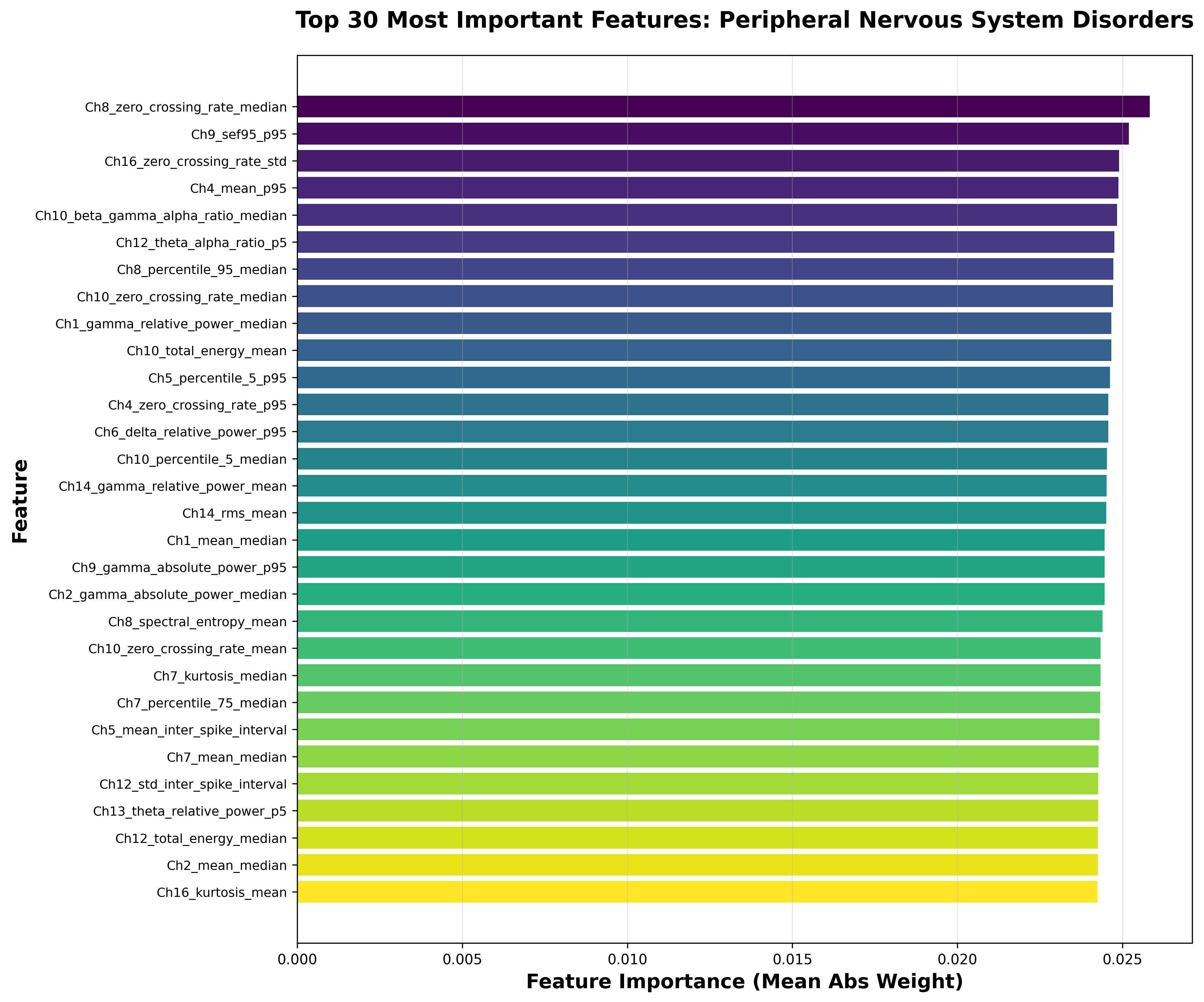}
\caption{Top 30 most informative features for Peripheral Nervous System Disorders. Frequency-domain and zero-crossing measures dominate, reflecting altered neuromuscular activity.}
\label{fig:feature_importance_peripheralns}
\end{figure}

Table~\ref{tab:peripheralns_top10_features} and Fig.~\ref{fig:feature_importance_peripheralns} summarize feature importance for Peripheral Nervous System Disorders.

\begin{table}[htbp]
\centering
\caption{Top ten most important features for Peripheral Nervous System Disorders based on MLP weight magnitude.}
\label{tab:peripheralns_top10_features}
\begin{tabular}{llc}
\hline
Rank & Feature & Importance \\
\hline
1 & Ch8\_zero\_crossing\_rate\_median & 0.0258 \\
2 & Ch9\_sef95\_p95 & 0.0252 \\
3 & Ch16\_zero\_crossing\_rate\_std & 0.0249 \\
4 & Ch4\_mean\_p95 & 0.0249 \\
5 & Ch10\_beta\_gamma\_alpha\_ratio\_median & 0.0248 \\
6 & Ch12\_theta\_alpha\_ratio\_p5 & 0.0248 \\
7 & Ch8\_percentile\_95\_median & 0.0247 \\
8 & Ch10\_zero\_crossing\_rate\_median & 0.0247 \\
9 & Ch1\_gamma\_relative\_power\_median & 0.0247 \\
10 & Ch10\_total\_energy\_mean & 0.0247 \\
\hline
\end{tabular}
\end{table}

\subsection{Impact of Threshold Optimization}

\begin{figure*}[htbp]
\centering
\begin{subfigure}{0.48\textwidth}
\includegraphics[width=\linewidth]{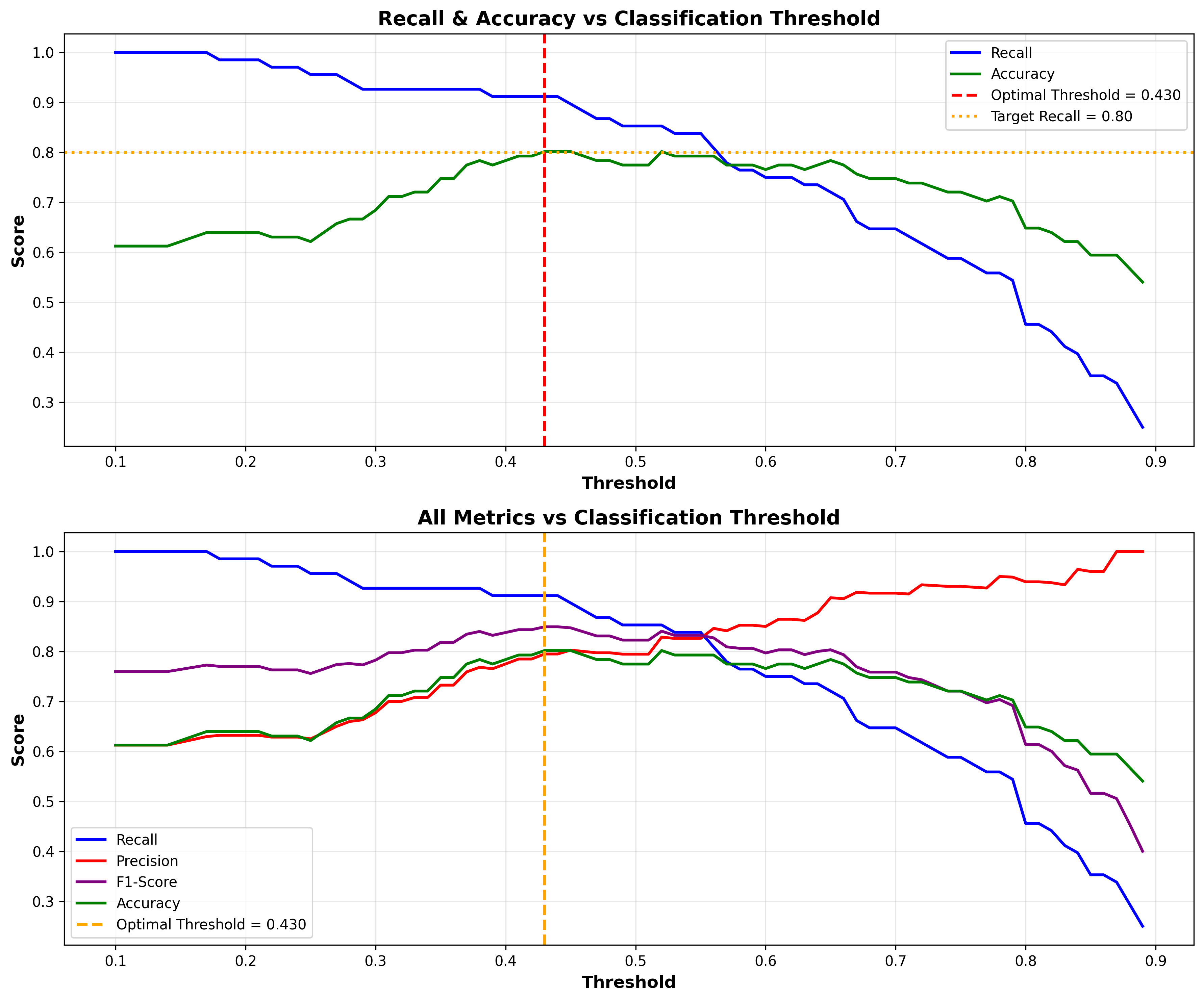}
\caption{Seizure Disorders (XGBoost).}
\end{subfigure}
\hfill
\begin{subfigure}{0.48\textwidth}
\includegraphics[width=\linewidth]{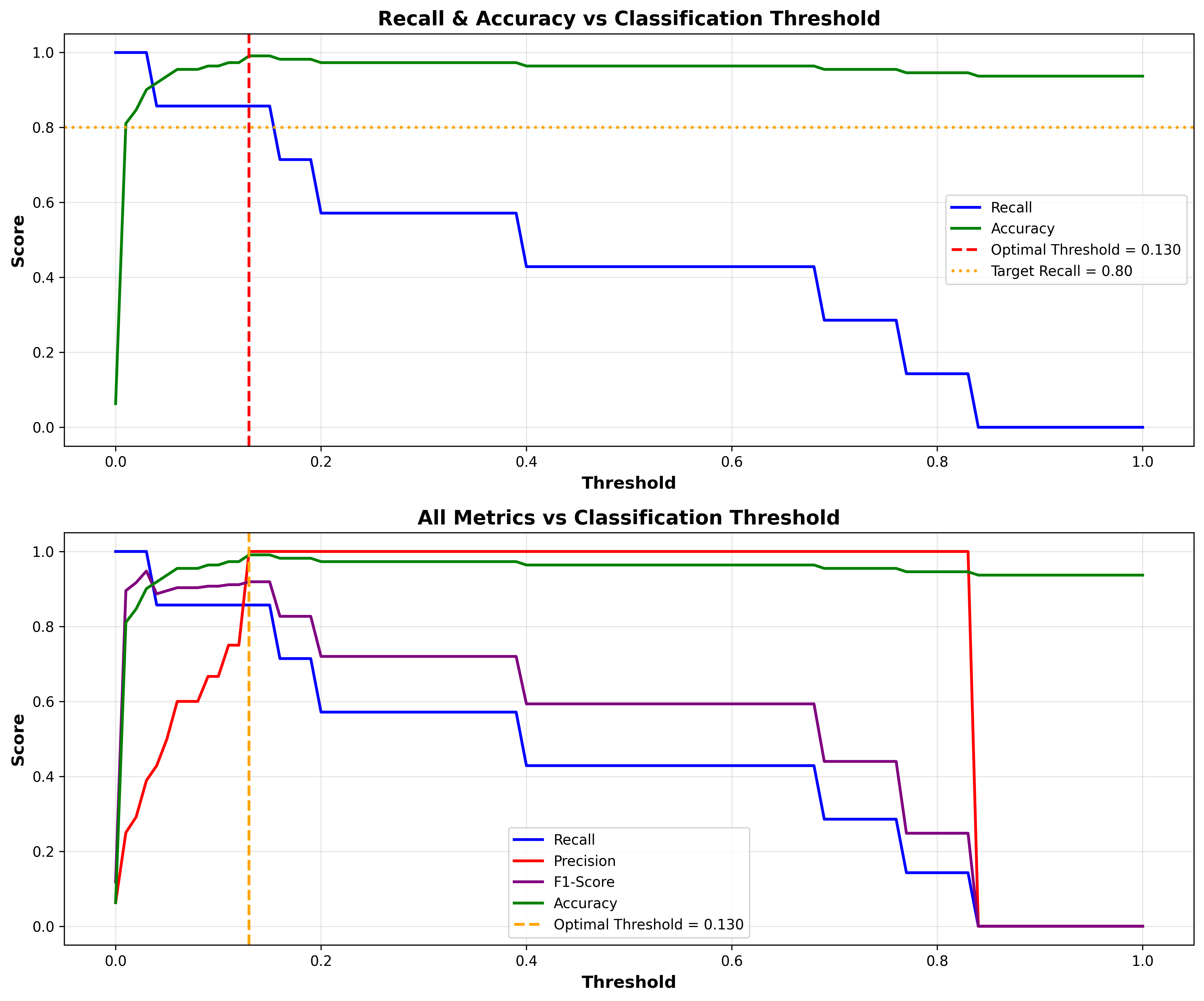}
\caption{Peripheral Nervous System Disorders (MLP).}
\end{subfigure}
\caption{Effect of decision threshold on recall, precision, accuracy, and F1-score. Optimized thresholds (dashed lines) achieve target recall while preserving accuracy.}
\label{fig:threshold_analysis}
\end{figure*}

Figure~\ref{fig:threshold_analysis} and Table~\ref{tab:threshold_impact} quantify the effect of decision threshold optimization on recall.

\begin{table}[htbp]
\centering
\caption{Absolute recall improvement achieved through threshold optimization.}
\label{tab:threshold_impact}
\begin{tabular}{lccc}
\hline
Disorder & Default & Optimized & Gain \\
\hline
Seizure Disorders & 85.3\% & 91.2\% & +5.9\% \\
Peripheral NS Disorders & 42.9\% & 85.7\% & +42.8\% \\
Developmental Delay & 33.3\% & 83.3\% & +50.0\% \\
\hline
\end{tabular}
\end{table}

\subsection{Summary of Key Results}

\begin{enumerate}
    \item Performance benchmarks are established across 11 neurological disorders using real-world clinical EEG data.
    \item All disorders achieve at least 75\% recall, with 10 of 11 meeting or exceeding the 80\% clinical sensitivity target.
    \item Model selection aligned with disorder characteristics improves robustness under severe class imbalance.
    \item Threshold optimization yields substantial recall gains (6--50\%) without unacceptable loss of accuracy.
    \item Feature importance patterns are physiologically interpretable and consistent with known EEG biomarkers.
\end{enumerate}

\section{Discussion}
\label{sec:discussion}

This study evaluates automated EEG-based classification across eleven clinically relevant neurological disorder categories using routine hospital EEG recordings. In contrast to task-specific EEG studies that focus on a single pathology—most commonly seizure detection or binary abnormality screening—this work addresses a clinically realistic scenario in which multiple diagnostic hypotheses must be considered concurrently from heterogeneous, noisy, and severely imbalanced data. This setting closely reflects real-world EEG usage, particularly for screening, triage, and longitudinal monitoring in hospital environments \cite{Benbadis2020,Tatum2014}.

The results demonstrate that clinically meaningful sensitivity can be achieved across diverse neurological conditions when model selection and decision calibration are aligned with electrophysiological characteristics and clinical priorities. Across all eleven disorders, recall exceeds 75\%, and for ten disorders meets or surpasses the 80\% sensitivity threshold commonly regarded as necessary for screening-oriented clinical decision support \cite{Benbadis2020,Halford2017}. Achieving this level of sensitivity across heterogeneous conditions is notable given the presence of severe class imbalance (up to 16.8:1) and the use of routine clinical EEG rather than curated research datasets.

These findings indicate that automated EEG analysis need not be restricted to narrowly defined tasks such as seizure detection to retain clinical utility. Instead, multi-disorder classification can operate as a high-sensitivity screening layer that flags recordings requiring expedited expert review. This framing is consistent with prior clinical guidance that emphasizes automation as a decision-support and triage mechanism rather than a replacement for expert interpretation \cite{Nuwer2018,Tatum2014}.

Clear performance differences between model architectures provide insight into disorder-dependent EEG structure. Gradient boosting models demonstrate stronger and more stable performance for seizure-related, rhythmic, and focal disorders, which are characterized by transient high-amplitude events, abrupt spectral shifts, and localized abnormalities. These features are well captured by tree-based ensembles operating on spectral and energy-domain biomarkers, consistent with prior seizure-focused EEG studies \cite{Shoeb2010,Lawhern2018}.

In contrast, Multi-Layer Perceptrons consistently outperform gradient boosting for peripheral, developmental, cognitive, and movement-related disorders. These conditions often manifest through subtle, spatially distributed, and temporally diffuse EEG alterations rather than discrete events. The superior performance of MLPs suggests that nonlinear interactions across multiple feature domains are required to capture these effects, aligning with observations from EEG studies in dementia, developmental disorders, and neuropsychiatric conditions \cite{Dauwels2010,Babiloni2016,Jeong2004}. Collectively, these results argue against a single unified classifier for all EEG-based diagnostic tasks and instead support a disorder-aware modeling strategy that mirrors clinical EEG interpretation practices.

Decision threshold optimization emerges as a central methodological component rather than a post-processing convenience. Absolute recall gains ranging from 6\% to 50\% demonstrate that default probability thresholds are insufficient in the presence of severe class imbalance and asymmetric clinical risk. Similar findings have been reported in other medical decision-support systems, where probability calibration and threshold selection substantially influence clinical utility \cite{Guo2017Calibration,Saito2015}. Explicitly incorporating threshold optimization aligns algorithmic behavior with clinical priorities, particularly in scenarios where false negatives carry disproportionate risk.

Feature importance analyses further support the physiological plausibility of the learned models. For seizure disorders, dominant features include high-frequency power variability, spectral edge frequency, entropy, and extreme amplitude measures—hallmarks of epileptiform activity and increased neuronal synchronization \cite{Fisher2014,Halford2017}. For peripheral nervous system disorders, the prominence of zero-crossing rates, spectral edge features, and power ratios is consistent with altered conduction dynamics and muscle-related activity that influence scalp EEG recordings \cite{Cassim2001}. This alignment between model-selected features and established neurophysiology reduces the likelihood that performance gains arise from dataset-specific artifacts and strengthens interpretability, an increasingly important consideration for clinical deployment and regulatory evaluation \cite{Holzinger2019}.

Several limitations merit consideration. First, the analysis is restricted to data from a single hospital site, and external validity across institutions, acquisition hardware, and patient populations remains to be demonstrated. Second, disorder labels are derived from ICD-10 diagnoses rather than EEG-specific expert consensus, introducing label noise that may underestimate achievable performance. Third, classification is performed at the recording level without temporal localization of pathological events. While sufficient for screening and triage, applications such as seizure onset localization or treatment monitoring would require finer temporal resolution. Finally, the reliance on handcrafted features prioritizes interpretability and robustness but may limit performance relative to end-to-end deep learning approaches when substantially larger annotated datasets become available.

Taken together, these findings support the feasibility of automated EEG-based screening across multiple neurological disorders using routine clinical recordings. By reducing diagnostic latency, standardizing preliminary assessment, and prioritizing expert attention toward high-risk cases, such systems address a growing clinical need in settings where EEG volume continues to increase while specialist availability remains limited \cite{Benbadis2020,Tatum2014}.

\section{Conclusion}
\label{sec:conclusion}
This work establishes a clinically grounded performance baseline for automated EEG-based classification across eleven neurological disorder categories using real-world hospital recordings from the Harvard Electroencephalography Database. By addressing multiple disorders simultaneously, the study moves beyond narrowly scoped EEG classification tasks and reflects the diagnostic complexity encountered in routine clinical practice.

The results demonstrate that clinically relevant sensitivity can be achieved across heterogeneous disorders when model architecture, feature representation, and decision thresholds are aligned with electrophysiological characteristics and clinical priorities. A hybrid modeling strategy combining gradient boosting and neural networks proves effective in capturing both focal, event-driven abnormalities and diffuse, nonlinear EEG alterations. Importantly, explicit threshold optimization is shown to be essential for maintaining high diagnostic sensitivity under severe class imbalance.

Feature importance analyses further reinforce the physiological validity of the framework, with dominant features corresponding to established EEG biomarkers for seizures, cognitive impairment, and peripheral nervous system pathology. This interpretability strengthens confidence in the reported performance and facilitates potential clinical translation.

While limitations related to single-site data, recording-level labels, and handcrafted features remain, these constraints define clear directions for future work, including multi-center validation, temporal localization of pathological activity, and integration of representation learning approaches. Overall, the findings support the role of automated EEG analysis as a screening and triage tool that complements expert interpretation and provides a scalable pathway toward more standardized and objective EEG-based decision support in clinical neurophysiology.

\bibliographystyle{IEEEtran}
\bibliography{references}

\begin{thebibliography}{10}
\providecommand{\url}[1]{#1}
\csname url@samestyle\endcsname
\providecommand{\newblock}{\relax}
\providecommand{\bibinfo}[2]{#2}
\providecommand{\BIBentrySTDinterwordspacing}{\spaceskip=0pt\relax}
\providecommand{\BIBentryALTinterwordstretchfactor}{4}
\providecommand{\BIBentryALTinterwordspacing}{\spaceskip=\fontdimen2\font plus
\BIBentryALTinterwordstretchfactor\fontdimen3\font minus \fontdimen4\font\relax}
\providecommand{\BIBforeignlanguage}[2]{{%
\expandafter\ifx\csname l@#1\endcsname\relax
\typeout{** WARNING: IEEEtran.bst: No hyphenation pattern has been}%
\typeout{** loaded for the language `#1'. Using the pattern for}%
\typeout{** the default language instead.}%
\else
\language=\csname l@#1\endcsname
\fi
#2}}
\providecommand{\BIBdecl}{\relax}
\BIBdecl

\bibitem{Tatum2014}
W.~O. Tatum, ``Handbook of eeg interpretation,'' \emph{Demos Medical Publishing}, 2014.

\bibitem{Westover2015}
M.~B. Westover, M.~M. Shafi, and M.~T. Bianchi, ``Continuous eeg monitoring in critically ill patients,'' \emph{Journal of Clinical Neurophysiology}, vol.~32, no.~2, pp. 87--95, 2015.

\bibitem{Schomer2018}
D.~L. Schomer and F.~L. da~Silva, \emph{Niedermeyer's Electroencephalography}.\hskip 1em plus 0.5em minus 0.4em\relax Oxford University Press, 2018.

\bibitem{Benbadis2008}
S.~R. Benbadis and K.~Lin, ``Errors in {EEG} interpretation and misdiagnosis of epilepsy. {W}hich {EEG} patterns are overread?'' \emph{European Neurology}, vol.~59, no.~5, pp. 267--271, 2008.

\bibitem{Hirsch2013}
L.~J. Hirsch, S.~M. LaRoche, N.~Gaspard \emph{et~al.}, ``Continuous eeg monitoring in the intensive care unit,'' \emph{Journal of Clinical Neurophysiology}, vol.~30, no.~1, pp. 1--27, 2013.

\bibitem{Halford2017}
J.~J. Halford, ``Interobserver agreement in eeg interpretation,'' \emph{Clinical Neurophysiology}, vol. 128, no.~3, pp. 513--520, 2017.

\bibitem{Shoeb2010}
A.~Shoeb and J.~Guttag, ``Application of machine learning to epileptic seizure detection,'' \emph{Proceedings of the 27th International Conference on Machine Learning}, 2010.

\bibitem{Roy2019}
Y.~Roy, H.~Banville \emph{et~al.}, ``Deep learning-based electroencephalography analysis,'' \emph{IEEE Transactions on Biomedical Engineering}, vol.~66, no.~9, pp. 2593--2604, 2019.

\bibitem{Jeong2004}
J.~C. Jeong, ``Eeg dynamics in patients with alzheimer's disease,'' \emph{Clinical Neurophysiology}, vol. 115, pp. 1490--1505, 2004.

\bibitem{Babiloni2016}
C.~Babiloni \emph{et~al.}, ``Abnormal cortical neural synchronization mechanisms in alzheimer's disease,'' \emph{Neurobiology of Aging}, vol.~40, pp. 95--105, 2016.

\bibitem{Feigin2019}
V.~L. Feigin \emph{et~al.}, ``Global, regional, and national burden of stroke, 1990--2016,'' \emph{The Lancet Neurology}, vol.~18, pp. 439--458, 2019.

\bibitem{WHO2021Dementia}
{World Health Organization}, ``Global status report on the public health response to dementia,'' 2021.

\bibitem{Dauwels2010}
J.~Dauwels \emph{et~al.}, ``Diagnosis of alzheimer's disease from eeg signals,'' \emph{Physiological Measurement}, vol.~31, pp. 537--560, 2010.

\bibitem{Nuwer2018}
M.~R. Nuwer, ``Assessment of digital eeg, quantitative eeg, and eeg brain mapping,'' \emph{Neurology}, vol.~49, no.~1, pp. 277--292, 1997.

\bibitem{Cassim2001}
F.~Cassim and W.~Szurhaj, ``Sedation and neuromuscular blockade: Effects on eeg and evoked potentials,'' \emph{Clinical Neurophysiology}, vol. 112, no.~6, pp. 1022--1032, 2001.

\bibitem{GBD2019Migraine}
{GBD 2019 Diseases and Injuries Collaborators}, ``Global, regional, and national burden of migraine,'' \emph{The Lancet Neurology}, vol.~19, pp. 954--976, 2020.

\bibitem{Fawcett2006}
T.~Fawcett, ``An introduction to roc analysis,'' \emph{Pattern Recognition Letters}, vol.~27, pp. 861--874, 2006.

\bibitem{Saito2015}
T.~Saito and M.~Rehmsmeier, ``The precision-recall plot is more informative than the roc plot,'' \emph{PLoS ONE}, vol.~10, no.~3, 2015.

\bibitem{Sun2025}
\BIBentryALTinterwordspacing
C.~Sun, J.~Jing, N.~Turley, C.~Alcott, W.~Y. Kang, A.~J. Cole, D.~M. Goldenholz, A.~Lam, E.~Amorim, C.~Chu, S.~Cash, V.~M. Junior, A.~Gupta, M.~Ghanta, B.~Nearing, F.~A. Nascimento, A.~Struck, J.~Kim, S.~Sartipi, A.~M. Tauton, M.~Fernandes, H.~Sun, G.~Bayas, K.~Gallagher, J.~B. Wagenaar, N.~Sinha, C.~Lee-Messer, C.~T. Silvers, B.~Gunapati, J.~Rosand, J.~Peters, T.~Loddenkemper, J.~W. Lee, S.~Zafar, and M.~B. Westover, ``Harvard electroencephalography database: A comprehensive clinical electroencephalographic resource from four boston hospitals,'' \emph{Epilepsia}, vol.~66, no.~9, pp. 3411--3425, September 2025. [Online]. Available: \url{https://doi.org/10.1111/epi.18487}
\BIBentrySTDinterwordspacing

\bibitem{Zafar2025}
\BIBentryALTinterwordspacing
S.~Zafar, T.~Loddenkemper, J.~W. Lee, A.~Cole, D.~Goldenholz, J.~Peters, A.~Lam, E.~Amorim, C.~Chu, S.~Cash, V.~M. Junior, A.~Gupta, M.~Ghanta, M.~Fernandes, H.~Sun, J.~Jing, and M.~B. Westover, ``Harvard electroencephalography database (version 4.1),'' Brain Data Science Platform, 2025. [Online]. Available: \url{https://doi.org/10.60508/k85b-fc87}
\BIBentrySTDinterwordspacing

\bibitem{article}
H.~Uyanik, A.~Sengur, M.~Salvi, R.~S. Tan, J.~H. Tan, and U.~Acharya, ``Automated detection of neurological and mental health disorders using eeg signals and artificial intelligence: A systematic review,'' \emph{Wiley Interdisciplinary Reviews: Data Mining and Knowledge Discovery}, vol.~15, 03 2025.

\bibitem{Rahul2024EEGSchizophreniaReview}
\BIBentryALTinterwordspacing
J.~Rahul, D.~Sharma, L.~D. Sharma, U.~Nanda, and A.~K. Sarkar, ``A systematic review of eeg-based automated schizophrenia classification through machine learning and deep learning,'' \emph{Frontiers in Human Neuroscience}, vol.~18, p. 1347082, February 2024. [Online]. Available: \url{https://doi.org/10.3389/fnhum.2024.1347082}
\BIBentrySTDinterwordspacing

\bibitem{10.1145/3742795}
\BIBentryALTinterwordspacing
A.~Belhadi, A.~Yazidi, P.~G. Lind, and Y.~Djenouri, ``Eeg data classification: Review and taxonomy,'' \emph{ACM Trans. Comput. Healthcare}, vol.~6, no.~4, Oct. 2025. [Online]. Available: \url{https://doi.org/10.1145/3742795}
\BIBentrySTDinterwordspacing

\bibitem{PARSA2023107683}
\BIBentryALTinterwordspacing
M.~Parsa, H.~Y. Rad, H.~Vaezi, G.-A. Hossein-Zadeh, S.~K. Setarehdan, R.~Rostami, H.~Rostami, and A.-H. Vahabie, ``Eeg-based classification of individuals with neuropsychiatric disorders using deep neural networks: A systematic review of current status and future directions,'' \emph{Computer Methods and Programs in Biomedicine}, vol. 240, p. 107683, 2023. [Online]. Available: \url{https://www.sciencedirect.com/science/article/pii/S0169260723003486}
\BIBentrySTDinterwordspacing

\bibitem{Shams2020NeurologicalDisorderDL}
\BIBentryALTinterwordspacing
M.~Shams and A.~Sagheer, ``A natural evolution optimization based deep learning algorithm for neurological disorder classification,'' \emph{Biomedical Materials and Engineering}, vol.~31, no.~2, pp. 73--94, 2020. [Online]. Available: \url{https://doi.org/10.3233/BME-201081}
\BIBentrySTDinterwordspacing

\bibitem{DONG2025108982}
\BIBentryALTinterwordspacing
C.~Dong, Z.~Zhang, and D.~Sun, ``Multi-channel eeg-based neurological disorder classification using cross-dependency spatiotemporal interactive network,'' \emph{Computer Methods and Programs in Biomedicine}, vol. 271, p. 108982, 2025. [Online]. Available: \url{https://www.sciencedirect.com/science/article/pii/S0169260725003992}
\BIBentrySTDinterwordspacing

\bibitem{GILL2024109732}
\BIBentryALTinterwordspacing
T.~S. Gill, S.~S.~H. Zaidi, and M.~A. Shirazi, ``Attention-based deep convolutional neural network for classification of generalized and focal epileptic seizures,'' \emph{Epilepsy \& Behavior}, vol. 155, p. 109732, 2024. [Online]. Available: \url{https://www.sciencedirect.com/science/article/pii/S1525505024001136}
\BIBentrySTDinterwordspacing

\bibitem{Amer2024VisualSignalNeuroscience}
\BIBentryALTinterwordspacing
N.~S. Amer and S.~B. Belhaouari, ``Exploring new horizons in neuroscience disease detection through innovative visual signal analysis,'' \emph{Scientific Reports}, vol.~14, no.~1, p. 4217, February 2024. [Online]. Available: \url{https://doi.org/10.1038/s41598-024-54416-y}
\BIBentrySTDinterwordspacing

\bibitem{Mathiyazhagan2025MotorImageryEEG}
\BIBentryALTinterwordspacing
S.~Mathiyazhagan and M.~S.~G. Devasena, ``Motor imagery eeg signal classification using novel deep learning algorithm,'' \emph{Scientific Reports}, vol.~15, p. 24539, 2025. [Online]. Available: \url{https://doi.org/10.1038/s41598-025-00824-7}
\BIBentrySTDinterwordspacing

\bibitem{ZEYNALI2023105130}
\BIBentryALTinterwordspacing
M.~Zeynali, H.~Seyedarabi, and R.~Afrouzian, ``Classification of eeg signals using transformer based deep learning and ensemble models,'' \emph{Biomedical Signal Processing and Control}, vol.~86, p. 105130, 2023. [Online]. Available: \url{https://www.sciencedirect.com/science/article/pii/S1746809423005633}
\BIBentrySTDinterwordspacing

\bibitem{Khamthung2024EmotionEEGXGBoost}
\BIBentryALTinterwordspacing
J.~Khamthung, N.~Lohia, and S.~Srivastava, ``Multi-class emotion classification with xgboost model using wearable eeg headband data,'' \emph{SMU Data Science Review}, vol.~8, no.~1, p. Article 7, 2024. [Online]. Available: \url{https://scholar.smu.edu/datasciencereview/vol8/iss1/7}
\BIBentrySTDinterwordspacing

\bibitem{9342132}
M.~Vijay, A.~Kashyap, A.~Nagarkatti, S.~Mohanty, R.~Mohan, and N.~Krupa, ``Extreme gradient boosting classification of motor imagery using common spatial patterns,'' in \emph{2020 IEEE 17th India Council International Conference (INDICON)}, 2020, pp. 1--5.

\bibitem{Khan2022CNNXGBoostEEG}
\BIBentryALTinterwordspacing
M.~S. Khan, N.~Salsabil, and M.~G.~R. Alam, ``Cnn-xgboost fusion-based affective state recognition using eeg spectrogram image analysis,'' \emph{Scientific Reports}, vol.~12, p. 14122, 2022. [Online]. Available: \url{https://doi.org/10.1038/s41598-022-18257-x}
\BIBentrySTDinterwordspacing

\bibitem{e22020140}
\BIBentryALTinterwordspacing
J.~Wu, T.~Zhou, and T.~Li, ``Detecting epileptic seizures in eeg signals with complementary ensemble empirical mode decomposition and extreme gradient boosting,'' \emph{Entropy}, vol.~22, no.~2, 2020. [Online]. Available: \url{https://www.mdpi.com/1099-4300/22/2/140}
\BIBentrySTDinterwordspacing

\bibitem{Vanabelle2019SeizureXGBoost}
\BIBentryALTinterwordspacing
P.~Vanabelle, P.~D. Handschutter, R.~E. Tahry, M.~Benjelloun, and M.~Boukhebouze, ``Epileptic seizure detection using eeg signals and extreme gradient boosting,'' \emph{Journal of Biomedical Research}, vol.~34, no.~3, pp. 228--239, August 2019. [Online]. Available: \url{https://doi.org/10.7555/JBR.33.20190016}
\BIBentrySTDinterwordspacing

\bibitem{WANG2022116778}
\BIBentryALTinterwordspacing
F.~Wang, Y.-C. Tian, X.~Zhang, and F.~Hu, ``An ensemble of xgboost models for detecting disorders of consciousness in brain injuries through eeg connectivity,'' \emph{Expert Systems with Applications}, vol. 198, p. 116778, 2022. [Online]. Available: \url{https://www.sciencedirect.com/science/article/pii/S095741742200238X}
\BIBentrySTDinterwordspacing

\bibitem{10.3389/fnins.2025.1541062}
\BIBentryALTinterwordspacing
A.~Belhadi, P.~G. Lind, Y.~Djenouri, and A.~Yazidi, ``Enhanced visibility graph for eeg classification,'' \emph{Frontiers in Neuroscience}, vol. Volume 19 - 2025, 2025. [Online]. Available: \url{https://www.frontiersin.org/journals/neuroscience/articles/10.3389/fnins.2025.1541062}
\BIBentrySTDinterwordspacing

\bibitem{SHARMA2022103101}
\BIBentryALTinterwordspacing
R.~Sharma, M.~Kim, and A.~Gupta, ``Motor imagery classification in brain-machine interface with machine learning algorithms: Classical approach to multi-layer perceptron model,'' \emph{Biomedical Signal Processing and Control}, vol.~71, p. 103101, 2022. [Online]. Available: \url{https://www.sciencedirect.com/science/article/pii/S1746809421006984}
\BIBentrySTDinterwordspacing

\bibitem{Hamzah2016EEGMotorMLP}
\BIBentryALTinterwordspacing
N.~Hamzah, H.~Norhazman, N.~Zaini, and M.~Sani, ``Classification of eeg signals based on different motor movement using multi-layer perceptron artificial neural network,'' \emph{Journal of Biological Sciences}, vol.~16, pp. 265--271, 2016. [Online]. Available: \url{https://scialert.net/abstract/?doi=jbs.2016.265.271}
\BIBentrySTDinterwordspacing

\bibitem{RAGHU2017205}
\BIBentryALTinterwordspacing
S.~Raghu and N.~Sriraam, ``Optimal configuration of multilayer perceptron neural network classifier for recognition of intracranial epileptic seizures,'' \emph{Expert Systems with Applications}, vol.~89, pp. 205--221, 2017. [Online]. Available: \url{https://www.sciencedirect.com/science/article/pii/S0957417417305067}
\BIBentrySTDinterwordspacing

\bibitem{4428831}
Y.-P. Lin, C.-H. Wang, T.-L. Wu, S.-K. Jeng, and J.-H. Chen, ``Multilayer perceptron for eeg signal classification during listening to emotional music,'' in \emph{TENCON 2007 - 2007 IEEE Region 10 Conference}, 2007, pp. 1--3.

\bibitem{9896584}
A.~Behboodi, W.~A. Lee, T.~C. Bulea, and D.~L. Damiano, ``Evaluation of multi-layer perceptron neural networks in predicting ankle dorsiflexion in healthy adults using movement-related cortical potentials for bci-neurofeedback applications,'' in \emph{2022 International Conference on Rehabilitation Robotics (ICORR)}, 2022, pp. 1--5.

\bibitem{10.3389/fnins.2016.00196}
\BIBentryALTinterwordspacing
I.~Obeid and J.~Picone, ``The temple university hospital eeg data corpus,'' \emph{Frontiers in Neuroscience}, vol. Volume 10 - 2016, 2016. [Online]. Available: \url{https://www.frontiersin.org/journals/neuroscience/articles/10.3389/fnins.2016.00196}
\BIBentrySTDinterwordspacing

\bibitem{Klem1999TenTwentySystem}
G.~H. Klem, H.~O. L{\"u}ders, H.~H. Jasper, and C.~Elger, ``The ten--twenty electrode system of the international federation,'' \emph{Electroencephalography and Clinical Neurophysiology Supplements}, vol.~52, pp. 3--6, 1999.

\bibitem{Baker2008EEGMCI}
\BIBentryALTinterwordspacing
M.~Baker, K.~Akrofi, R.~Schiffer, and M.~W. Boyle, ``Eeg patterns in mild cognitive impairment (mci) patients,'' \emph{The Open Neuroimaging Journal}, vol.~2, pp. 52--55, 2008. [Online]. Available: \url{https://doi.org/10.2174/1874440000802010052}
\BIBentrySTDinterwordspacing

\bibitem{hjorth1970eeg}
B.~Hjorth, ``Eeg analysis based on time domain properties,'' \emph{Electroencephalography and clinical neurophysiology}, vol.~29, no.~3, pp. 306--310, 1970.

\bibitem{6773024}
C.~E. Shannon, ``A mathematical theory of communication,'' \emph{The Bell System Technical Journal}, vol.~27, no.~3, pp. 379--423, 1948.

\bibitem{Welch1967PowerSpectra}
P.~D. Welch, ``The use of fast fourier transform for the estimation of power spectra: A method based on time averaging over short, modified periodograms,'' \emph{IEEE Transactions on Audio and Electroacoustics}, vol. AU-15, no.~2, pp. 70--73, June 1967.

\bibitem{10.1098/rspl.1895.0041}
\BIBentryALTinterwordspacing
K.~Pearson, ``Vii. note on regression and inheritance in the case of two parents,'' \emph{Proceedings of the Royal Society of London}, vol.~58, no. 347-352, pp. 240--242, 12 1895. [Online]. Available: \url{https://doi.org/10.1098/rspl.1895.0041}
\BIBentrySTDinterwordspacing

\bibitem{Pearson01111901}
\BIBentryALTinterwordspacing
------, ``Liii. on lines and planes of closest fit to systems of points in space,'' \emph{The London, Edinburgh, and Dublin Philosophical Magazine and Journal of Science}, vol.~2, no.~11, pp. 559--572, 1901. [Online]. Available: \url{https://doi.org/10.1080/14786440109462720}
\BIBentrySTDinterwordspacing

\bibitem{Chen_2016}
\BIBentryALTinterwordspacing
T.~Chen and C.~Guestrin, ``Xgboost: A scalable tree boosting system,'' in \emph{Proceedings of the 22nd ACM SIGKDD International Conference on Knowledge Discovery and Data Mining}, ser. KDD ’16.\hskip 1em plus 0.5em minus 0.4em\relax ACM, Aug. 2016, p. 785–794. [Online]. Available: \url{http://dx.doi.org/10.1145/2939672.2939785}
\BIBentrySTDinterwordspacing

\bibitem{10.5555/3104322.3104425}
V.~Nair and G.~E. Hinton, ``Rectified linear units improve restricted boltzmann machines,'' in \emph{Proceedings of the 27th International Conference on International Conference on Machine Learning}, ser. ICML'10.\hskip 1em plus 0.5em minus 0.4em\relax Madison, WI, USA: Omnipress, 2010, p. 807–814.

\bibitem{kingma2017adammethodstochasticoptimization}
\BIBentryALTinterwordspacing
D.~P. Kingma and J.~Ba, ``Adam: A method for stochastic optimization,'' 2017. [Online]. Available: \url{https://arxiv.org/abs/1412.6980}
\BIBentrySTDinterwordspacing

\bibitem{Benbadis2020}
S.~R. Benbadis, ``The human error in eeg interpretation,'' \emph{Epilepsy \& Behavior}, vol. 111, p. 107019, 2020.

\bibitem{Lawhern2018}
V.~J. Lawhern, A.~J. Solon, N.~R. Waytowich, S.~M. Gordon, C.~P. Hung, and B.~J. Lance, ``Eegnet: A compact convolutional neural network for eeg-based brain--computer interfaces,'' \emph{Journal of Neural Engineering}, vol.~15, no.~5, p. 056013, 2018.

\bibitem{Guo2017Calibration}
C.~Guo, G.~Pleiss, Y.~Sun, and K.~Q. Weinberger, ``On calibration of modern neural networks,'' \emph{Proceedings of the 34th International Conference on Machine Learning}, vol.~70, pp. 1321--1330, 2017.

\bibitem{Fisher2014}
R.~S. Fisher, C.~Acevedo, A.~Arzimanoglou, and et~al., ``Ilae official report: A practical clinical definition of epilepsy,'' \emph{Epilepsia}, vol.~55, no.~4, pp. 475--482, 2014.

\bibitem{Holzinger2019}
A.~Holzinger, G.~Langs, H.~Denk, K.~Zatloukal, and H.~M{\"u}ller, ``Causability and explainability of artificial intelligence in medicine,'' \emph{WIREs Data Mining and Knowledge Discovery}, vol.~9, no.~4, p. e1312, 2019.

\end{thebibliography}

\end{document}